\PassOptionsToPackage{dvipsnames}{xcolor}

\documentclass[11pt]{article}
\usepackage{epsfig,endnotes,cite}
\usepackage{fullpage}
\usepackage{times}

\date{}

\newcommand{\htor}[1]{\noindent\textcolor{green}{[htor: #1]}}

\usepackage{amsmath,amssymb}

\usepackage{algorithm}
\usepackage{algorithmic}

\usepackage{footnote}
\usepackage{hyperref}

\newcommand{\R}{\mathbb{R}}

\newcommand{\normsmall}[1]{\| #1 \|}

\usepackage{booktabs}
\usepackage{subcaption}
\usepackage{listings}
\usepackage{amsthm}
\usepackage{bm}
\usepackage{enumitem}
\setitemize{noitemsep,topsep=0pt,parsep=0pt,partopsep=0pt}

\usepackage{tikz}
\usepackage{enumitem}
\usepackage{multirow}
\usetikzlibrary{positioning, fit, shapes, patterns, calc}

\renewcommand{\paragraph}[1]{\vspace{0.2em} \noindent \textit{\textbf{#1}}}

\newtheorem{theorem}{Theorem}[section]

\newtheorem{lemma}[theorem]{Lemma}

\newtheorem{definition}[theorem]{Definition}
\newtheorem{assumption}[theorem]{Assumption}

\newcommand{\mml}{$\textsc{SparCML}$}

\begin{document}
\title{\mml{}: High-Performance Sparse Communication \\ for Machine Learning}

\author{
{\rm Cedric Renggli}\\
ETH Zurich
\and
{\rm Saleh Ashkboos}\\
IST Austria
\and 
{\rm Mehdi Aghagolzadeh}\\
Microsoft
\and 
{\rm Dan Alistarh}\\
IST Austria
\and 
{\rm Torsten Hoefler}\\
ETH Zurich
} 

\maketitle

\begin{abstract}
\noindent
  Applying machine learning techniques to the quickly growing data in
  science and industry requires highly-scalable algorithms. 
  %
  %
  Large datasets are most commonly processed ``data parallel''
  distributed across many nodes. Each node's contribution to the
  overall gradient is summed using a global allreduce. 
  This allreduce is the single communication and thus scalability
  bottleneck for most machine learning workloads. 
  We observe that frequently, many gradient values are (close to) zero,
  leading to sparse of sparsifyable communications. 
  To exploit this insight, we analyze, design, and implement a set of
  communication-efficient protocols for sparse input data, in
  conjunction with efficient machine learning algorithms which can
  leverage these primitives. 
  %
  %
  %
  Our communication protocols generalize standard collective operations,
  by allowing processes to contribute \emph{arbitrary} sparse input data
  vectors.
  %
  %
  Our generic communication library, \mml{}\footnote{Stands for Sparse Communication layer for Machine Learning, to be read as \emph{sparse ML}.}, extends MPI to support
  additional features, such as non-blocking (asynchronous) operations
  and low-precision data representations. 
  As such, \mml{} and its techniques will form the basis of future
  highly-scalable machine learning frameworks.
 \end{abstract}

\section{Introduction and Motivation}

Machine learning workloads are quickly becoming more demanding. The
size of the trained models and with it the computation required for
training grow with the
current exponential growth of data availability. While small models used
to train in minutes on laptops, and image recognition networks such as
AlexNet required days on a GPU system, newer models such as
BERT~\cite{DBLP:journals/corr/abs-1810-04805}
would take more than one year to train on a single GPU~\cite{BERTtime}.
Similarly, the sizes of the networks grow quickly from a handful of
parameters for simple regression tasks to more than 200 MB for Alexnet
to up to 340 million parameters, i.e., 11 GB with 32 bit precision, for
the largest BERT network. 

The arguably standard distribution strategy in machine learning is
\emph{data parallelism}, in which nodes partition the dataset, and
maintain consistent copies of the set of model parameters computing a
global sum, either with allreduce, or through a coordinator node, called
a parameter server~\cite{PS}. Here, we consider only allreduce due
to the obvious scaling limitations of a parameter server.
While it's relatively simple to scale the number of execution nodes to
the thousands, the biggest bottleneck is the allreduce of the gradient
values at each step. The size of this reduction is equivalent to the
model size itself and it is not reduced when more nodes are used. 
When scaling to large numbers of nodes, the full parameter set, commonly
hundreds of megabytes, must be summed globally every few microseconds. 

%

Given the large impact of communication, significant effort has
been invested into identifying scalable solutions. Virtually all major
frameworks  optimize for efficient communication~\cite{CNTK, seide20141,
TF, MXNET, PyTorch}, while GPU vendors are developing specific
communication layers for this goal~\cite{NCCL}.  The research community
proposed several communication-reduction techniques, such as
\emph{quantization}~\cite{seide20141, strom2015scalable, alistarh2016qsgd},
\emph{asynchronous communication}~\cite{zhang2017yellowfin},
\emph{structured sparsification}~\cite{aji2017sparse,
dryden2016communication, sun2017meprop}, or \emph{large batch
methods}~\cite{goyal2017accurate, you2017scaling}. However, scaling
machine learning applications remains a complex process, requiring
non-trivial insights. In our work, we focus on a largely unexplored
aspect of scaling: \emph{how to exploit sparsity in the global summation
itself}. Our techniques build on established HPC techniques such as MPI
and extend it for generic support or arbitrary sparse reductions. 

\paragraph{Conceptual Contribution.} We propose \mml{}, a scalable,
general communication library for machine learning.  \mml{} starts from
the idea that, to reduce communication and synchronization cost, we can
exploit \emph{sparsity} and \emph{relaxed consistency} in machine learning
applications.  In particular, individual nodes can compute with a
partially \emph{inconsistent} view of the parameters. The immediate
system implication, which we exploit in \mml{}, is that the updates
which nodes wish to communicate are either \emph{naturally
sparse}~\cite{webb2006introducing}, or can be \emph{sparsified in a
principled manner}, without loss of convergence~\cite{alistarh2016qsgd,
aji2017sparse, dryden2016communication, sun2017meprop}. 

%
\paragraph{Technical Contribution.} Our thesis is that \emph{exploiting
sparsity and compression} should be standard when scaling machine
learning applications.  Surprisingly, support for efficient sparse
communication or compression is currently neither available in standard
communication libraries such as MPI~\cite{mpi-3.0}, nor in specialized
machine-learning communication libraries~\cite{NCCL}.  One possible
reason is the fact that designing and implementing general sparse
collective operations is non-trivial, as sparsity adds a new dimension
to the already complex system trade-offs arising when implementing
collective operations efficiently at scale~\cite{thakur2003improving}. 

We take on this challenge in \mml{}. 
Our implementation is efficient both in theory and in practice: for some workload parameters, it can be shown to be within constant factors of optimal in terms of bandwidth and latency cost. At the same time, our implementation achieves order-of-magnitude speedups versus highly optimized \emph{dense collective} implementations, or over naive sparse implementations, both in synthetic tests and in real application scenarios.  \mml{} has several additional features, such as efficient support for \emph{reduced-precision collectives} and for \emph{non-blocking} operations. 
For example, we can perform sparse reductions for gradient exchange at 4 bits of precision per coordinate, overlapping computation and communication. 

\paragraph{Targets.} Our main target applications are two large-scale distributed machine learning tasks: training of state-of-the-art deep neural networks and large-scale regularized classification tasks. 
Our target systems are multi-node computing clusters. We study two scenarios: the first is \emph{supercomputing}, where nodes are connected by a high-powered, extremely well optimized network.
The second scenario is \emph{datacenters}, where the network is \emph{relatively} slower, such as InfiniBand or Gigabit Ethernet.

\paragraph{Challenges.} The main algorithmic contribution behind our layer is a set of techniques for implementing collective communication operations, such as allreduce sum, over a large number of nodes having input vectors that are \emph{sparse}. 
The principal difficulty for designing and analyzing such algorithms lies in the unknown overlap of non-zero indices, and hence the size of the reduced result. 
We provide an adaptive set of techniques which can systematically handle all cases and their trade-offs.
These algorithmic insights are backed by careful optimizations and additional system features. 
An additional challenge from the machine learning side comes with avoiding additional  hyperparameter tuning in order to leverage sparsity---in our experiments, we find that this is possible, with a few notable exceptions.

\paragraph{Experimental Results.} 
\sloppy
We validate \mml{} on a range of benchmarks: 1) synthetic instances aimed to validate our analysis, 2) academic benchmark datasets and models, and 3) large-scale deployments for image classification and automated speech recognition (ASR). 
Synthetic benchmarks show that \mml{} can bring order-of-magnitude speedups with respect to highly-optimized dense implementations, with limited overhead in the dense case. 
We incorporate \mml{} into two machine learning frameworks: CNTK (developed by Microsoft) and MPI-OPT (developed by us). In the supercomputing deployment, \mml{} can reduce end-to-end convergence time of a state-of-the-art network for natural language understanding by $6\times$. Further, it completes a large-scale URL classification task $31\times$ faster than its Cray MPI-based variant, which however does not exploit sparsity. 
The speedups are more significant on less performant cloud networks. 

For large-scale workloads, we investigated training CNNs on the ImageNet dataset~\cite{russakovsky2015imagenet}, and training the LSTM networks powering the ASR component of a popular personal assistant. 
In the first scenario, we found that~\mml{} was able to reduce the end-to-end training time for wide residual networks~\cite{zagoruyko2016wide} on ImageNet by $\simeq 2\times$ on $64$ GPUs, with relatively negligible accuracy loss ($<0.5\%$ Top5 validation), and no additional hyperparameter tuning. However, gains were negligible when applied to the standard ResNet50 benchmark~\cite{he2016deep, goyal2017accurate}, which has fewer parameters and is therefore less amenable to training via sparse gradients. 
In the ASR task,~\mml{} reduced the training time for a state-of-the-art LSTM model on 128 GPUs by almost $10\times$ (from $14$ days to $1.78$ days), without significant accuracy loss.
Our conclusion is that \mml{} can yield non-trivial speedups on a variety of machine learning applications, 
and that existing frameworks can significantly leverage sparsity and relaxed consistency guarantees.

\vspace{-1em}
\section{Preliminaries}
\label{sec:prelim}


\paragraph{Notation.} Throughout this paper, we use the following notation for input parameters:
\begin{footnotesize}
	\begin{center}
		\begin{tabular}{ @{} c | l @{} }
			\toprule
			Variable & Description \\
			\midrule 
			$P$ & Number of nodes \\
			$N$ & Problem dimension \\
			$p_i$ & Node $i$, $1 \leq i \leq P$ \\
			$H_i$ & Set of non-zero indices which $p_i$ wishes to communicate \\
			$k$ & Max number of non-zero (nnz) elements: $\max_i \vert H_i \vert$ \\
			$\mathcal{K}$ &  Total nnz in global sum: $\vert \cup_{i=1}^{P} H_i \vert$ \\
			$d$ & Density of non-zero elements: $\frac{k}{N}$ \\
			\bottomrule
		\end{tabular}
	\end{center}
\end{footnotesize}

\subsection{Data Parallelism and Communication Costs}

\emph{Data-parallelism} is a standard distribution strategy for machine
learning algorithms~\cite{TF,
CNTK}: $P$ computing nodes share a large dataset and each maintains
its own copy of the model $\vec{x}_t$. Model copies are kept in sync across nodes by
exchanging the model updates computed locally between nodes, either via global
averaging of updates, or through a central coordinator~\cite{PS}.  
Specifically, in Stochastic Gradient Descent (SGD), each node $i$ has access to (part of) the dataset, and, in each \emph{iteration}, it processes a randomly chosen set of samples (a \emph{mini-batch}), and computes
a model update (gradient) $\nabla F_i(\vec{x}_t)$ locally. Nodes then globally \emph{sum} these updates, and apply them locally, resulting in the following standard SGD iteration 
$$
 \vec{x}_{t+1} = \vec{x}_t - \eta  \sum_{i = 1}^P \nabla F_i(\vec{x}_t),
$$
\noindent where $\vec{x}_t$ is the value of the model at time $t$,
$\eta$ is the learning rate, and $\nabla F_i$ is the \emph{stochastic}
gradient of the current model with respect to the set of samples processed at node $i$.\footnote{For simplicity, the reader may think of the model $\vec{x}_t$ as a large array of parameters, and of the gradients $\nabla F_i(\vec{x}_t)$ as array of entry-wise updates.}
Since gradient updates are averaged globally at the end of every
iteration, all nodes have a consistent version of the model.  The
trade-off is between the parallelism due to the fact
that we are processing $P$ times more samples per iteration given $P$
nodes, and the additional communication cost due to the sum reduction, necessary to maintain a consistent model. 
To reduce this overhead, several communication reduction techniques have been proposed. 
 
\subsection{Communication-Reduction Techniques}

\paragraph{Structured Sparsification.} Recent work proposes the following communication-reduced SGD variant
which we call Top-$k$ SGD~\cite{dryden2016communication,
aji2017sparse}: each node communicates only the $k$ largest (by
magnitude) components of its gradient vector $\nabla F(\vec{x}_t)$,
instead of all values in the traditional method. 
Usually, $k$ is fixed to represent some percentage of the components,
which can be even lower than $1\%$~\cite{lin2017deep}.
This forces gradient sparsity at each node, although the chosen components may
vary across nodes. The value of the components which are not chosen is \emph{accumulated}, and added to the gradient vector of the next iteration. A precise description of this procedure can be obtained by following Algorithm~\ref{algo:qtopk-sgd}, where the quantization function $Q$ should be taken to be the identity.

\begin{algorithm}[t!]
   \caption{\mml{} Quantized TopK SGD at a node $i$.}
   \label{algo:qtopk-sgd}
\begin{algorithmic}
   \STATE {\bfseries Input:} Stochastic Gradient $\nabla F_i(\cdot)$ at node $i$
   \STATE {\bfseries Input:} value $K$, learning rate $\alpha$
   \STATE  Initialize $v_0 = \epsilon_0^i = \vec{0}$
   \FOR{ each step $t \geq 1$}
    \STATE ${acc}_t^i \gets \epsilon_{t - 1}^i + \alpha \nabla F_i(v_{t - 1}) $ \COMMENT{accumulate error into a locally generated gradient}
    \STATE $\epsilon_t^i \gets {acc}_t^i -  \mathsf{TopK}( {acc}_t^i ) $ \COMMENT{update the error}
    \STATE $g_t^i \gets \mathsf{allreduce}( Q( \mathsf{TopK}( {acc}_t^i ) ), \mathsf{SUM}
    )$ \COMMENT{ sum (sparse) contribution from all nodes }
    \STATE  $v_{t}^i \gets v_{t-1}^i - g_t^i$  \COMMENT{ apply the update }
   \ENDFOR
\end{algorithmic}
\end{algorithm}

\paragraph{Quantization.} An orthogonal approach for reducing
the communication cost of machine learning algorithms has been to
\emph{quantize} their updates, lowering the number of bits used to
represent each value, e.g.~\cite{seide20141, de2015taming, alistarh2016qsgd, terngrad}. 
Mathematically, the resulting iteration can be represented as:

$$
 \vec{x}_{t+1} = \vec{x}_t - \eta  \sum_{i = 1}^P Q(\nabla F_i(\vec{x}_t)),
$$

\noindent where $Q: \R^N \rightarrow \R^N$ is an element-wise quantization operator which reduces the precision of the gradients' data representation. 
Quantization techniques can also be shown to preserve convergence, as
long as the quantization noise is zero-mean, but may slow down convergence due to added variance~\cite{alistarh2016qsgd}. 

\vspace{-1em}
\section{Communication-Reduction: A Critical View}

We now examine communication-reduction techniques in the context of large-scale deployments characteristic to super-computing or large-scale cloud computing.

\paragraph{Structured Sparsification.} On the positive side, sparsification has been proven to preserve convergence even for non-convex objectives~\cite{alistarh2018convergence}, and have been empirically shown, in the context of neural network training, to be able to allow nodes to send even less than $1\%$ of their local gradient update, without losing convergence~\cite{lin2017deep}. 

\begin{figure}[tbp]
	\centering
	\includegraphics[width=0.45\textwidth]{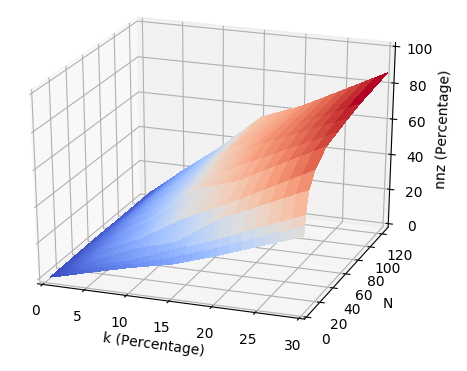}
	\caption{The density (in percentage) of the reduced result versus number of nodes $N$ and per-node density $k$ (in percentage) for the ResNet20 model trained on the CIFAR-10 dataset. The snapshot is taken at training epoch 5. Results are consistent across training stages (1st, 5th, and last epoch) and models (ResNet, DenseNet).}
	\label{fig:reduction-density}
\end{figure}

Unfortunately however, reaching high sparsity levels (above $99\%$) requires extremely careful tuning of momentum and learning rate hyperparameters, which is error-prone and time consuming. Employing \emph{lower} sparsity levels--say, $5-10\%$ per node, which tends to be more stable with respect to hyper-parameter tuning--can negate the benefits of compression: in this case, reducing across a large number of nodes cans cause the reduced vector to become \emph{dense}, at which point communication is again a bottleneck. We illustrate this issue in Figure~\ref{fig:reduction-density}, where we plot the density of the reduced gradient versus the number of nodes and the sparsity level at each node, on a standard CNN/dataset combination. 

\paragraph{Quantization.} These techniques do not have the issue that communication-compression is dependent of the node count:  their compression rate is fixed and independent of the number of nodes. However, both theory and experiments suggest that quantization can only yield a limited amount of compression ($4-8\times$) before the added variance affects end accuracy~\cite{alistarh2016qsgd, EDBT}.  

\section{Communication Reduction in \mml{}} 

In sum, the previous section suggests that neither sparsification nor quantization is ideal in isolation, when considered at high node counts. In this context, the \mml{} framework allows the user
to leverage both quantization and sparsification methods. We
provide efficient implementations of structured sparse methods (TopK SGD) via sparse
collective operations,  with non-blocking semantics, as well as an
implementation of state-of-the-art quantization
methods~\cite{alistarh2016qsgd}. 
Importantly, the implementation of the reduction method natively supports both of these techniques to reduce communication and latency constraints. 

\paragraph{Sparse Quantized Reduction.} We now provide a high-level algorithmic description of the data-parallel SGD variant implemented in \mml{}. Please see Algorithm~1 for pseudocode. 
Each node $i$ maintains the residual error $\epsilon^i$ locally, which accumulates gradient components which did not get applied. Upon each step $t$, this gets added to the newly generated gradient, to obtain the accumulator $acc_t^i$. 
This accumulator is then truncated to obtain the value $g_t^i$ to be sent by node $i$, and the new value of the error $\epsilon^i$ is generated. 

The allreduce call sums all the truncated gradients in a sparsity-aware fashion. In particular, since the sum may become \emph{dense}, \mml{} may quantize the resulting dense vector at an intermediate stage of the reduction, in order to reduce the bandwidth overhead, using stochastic QSGD quantization~\cite{alistarh2016qsgd}. 
We note that, even though sparsification and stochastic quantization have been introduced independently in the literature (see references~\cite{dryden2016communication, aji2017sparse, lin2017deep} and~\cite{seide20141, alistarh2016qsgd}, respectively), we are the first to employ these two techniques in conjunction, and to prove that the resulting method provably converges.   

\paragraph{Convergence Proof.} 
The following result formally states the convergence guarantees of \mml{}, under standard analytic assumptions on the objective function. The argument (provided in the appendix) is based on the convergence proof of TopK SGD~\cite{alistarh2018convergence}; the main source of novelty is the addition of stochastic quantization. 

\begin{theorem}
\label{theorem_nonconvex}
Consider the \mml{} SGD algorithm when minimising a smooth, non-convex function $f$. 
Then there exists a learning rate schedule $(\alpha_t)_{t = 1, T}$ such that the following  holds:

$$ \min_{t\in\{1, 2, \ldots, T\}} \mathbb{E}  \, [ \|\nabla f\left(x_t\right)\|^2 ]  \stackrel{T\rightarrow \infty}{\rightarrow} 0.$$
\end{theorem}

\paragraph{Discussion.} The above statement is quite general in that it covers a large class of non-convex objectives. However, it only proves \emph{ergodic convergence} to a stationary point of expected zero gradient. 
This is weaker than proving convergence to a global minimum, but is in line with state-of-the-art results for this problem setting even without quantization, e.g.~\cite{lian2015asynchronous}.  
A second limitation shared by most theoretical results is that it does not provide a precise set of hyperparameters for practical deployments, beyond the indication that learning rates should be diminishing.

\vspace{-1em}
\section{Supporting Sparsity in \mml{}}

\subsection{Data Representation: Sparse Streams}

We now describe the data types used to store sparse and
dense vectors, which we call \emph{sparse streams}. Sparse streams allow
for efficient computation and communication of the data. Our
implementation  is in \verb|C++11|, and we follow this standard in our
description. For simplicity, we focus on the case where the binary
operation executed upon two or multiple streams is \emph{summation}, but the same discussion would apply for other component-wise operations. 

Initially, we assume that each node is assigned a subset of non-zero
elements from a universe of size $N$. Let $H_i$ denote the set of
non-zero elements given at node $p_i$. We assume that these sets are
\emph{sparse} with respect to $N$, i.e., that $k = \max_i \vert H_i
\vert \ll N$. We further denote by $d_i$ the density of each set given
by $d_i = \frac{\vert H_i \vert}{N}$ and define $d = \max_{i} d_i =
\frac{k}{N}$.

We define the total number of non-zero elements after having performed the
reduction as $$\mathcal{K} = \vert \cup_{i=1}^P H_i \vert.$$ For
simplicity, we ignore  cancellation of
indices during the summation and therefore get $k \leq \mathcal{K} \leq \min \{N, P \times k\}.$ 

\paragraph{Vector Representations.}
We start from the standard sparse representation, storing a sparse vector as a
sequence of non-zero indices, together with the actual scalar values of
each dimension.  
The
stream is stored in an array of consecutive index-value pairs.
The datatype of the values yields the number of bits needed
for every non-zero value. We either work with single or
double precision floating point values. We discuss lower
precision support in Section~\ref{sec:additional-features}.  

\paragraph{Switching to a Dense Format.}
Although we are
interested in \emph{sparse} problems, the size and
non-zero index distribution of the input vectors can be such that the
algorithm may not benefit from the sparse representation after some
intermediate point in the summation process: as the density of the
intermediate result vector approaches the universe size $N$, the sparse
representation becomes wasteful. 

We can model the benefits of sparsity as follows: Let $isize$ be the
number of bytes needed to represent a non-zero input value and $nnz$ the
number of non-zero elements. We further define $c\geq \left\lceil
\frac{\log_2(N)}{8} \right\rceil$ to be the number of bytes needed to
store an index. 
Thus, the sparse format will transmit $nnz (c+isize)$ bytes while the
dense format transmits $N\times isize$ bytes.
Our sparse representation only reduces the communication volume if $nnz \leq
\delta = \frac{N \times isize}{(c+isize)}$. Yet, this volume estimation does
not capture the fact that summing sparse vectors is computationally more
expensive than summing dense vectors. Thus, in practice, $\delta$ should
be even smaller, to reflect this trade-off.

It is safe to assume that the initial $nnz=k$ is smaller than this
threshold. However, as the summation advances and number of nonzero
elements $nnz$ in the vector grows, the condition $nnz \leq \delta$ may be
violated. Especially for large node counts $P$, $\mathcal{K}$ is almost
certainly larger than $\delta$. 
To address this dynamic fill-in, we add an extra value to the beginning
of each vector that indicates whether the vector is dense or sparse.
In fact, when allocating memory for vectors of dimension $N$, we request
$N \times isize$ bytes. It is therefore never possible to store more
than $\delta$ sparse items. This threshold is used to automatically
switch the representation.  

\paragraph{Efficient Summation.}
The key operation is summing up two vectors $u_1$ and $u_2$, which could be
either sparse or dense. To implement this operation efficiently, we
distinguish two cases: (1) $u_1$ and $u_2$'s indices draw from any
position between $1$ and $N$, and can potentially overlap and (2) $u_1$
and $u_2$'s index sets are disjoint. The latter which arises if we
partition the problem by dimension in which case we can implement the
sum as simple concatenation.

If input indices can overlap, we distinguish the following cases depending on whether inputs are sparse or dense. 
Denote by $H_1$ and $H_2$ the sets containing the sparse indices of
non-zero elements for the vectors $u_1$ and $u_2$, respectively. 
%
%
%
%
%
%
%
%
If indices are overlapping, and both vectors
are sparse, we first check whether the result might become
dense.  Theoretically, one needs to calculate the size of the union of
non-zero indices $\vert H_1 \cup H_2 \vert$. 
This is costly, and thus we only upper bound this result by $\vert H_1 \vert + \vert H_2 \vert$. 
The tightness of this upper bound will depend on the underlying sparsity distribution, on which we make no prior assumptions.
If this value is bigger than
$\delta$, we switch to a dense representation. 
If one of the inputs is dense, whereas the other is sparse, we
iterate over all the index-value pairs stored in the sparse vector and
set the value at the corresponding position in the dense vector. Finally, if both vectors are already dense, we simply perform a (vectorized) dense vector summation in either $u_1$ or $u_2$, and do not allocate a new stream. 

\subsection{Efficient Collectives on Sparse Streams}

We now proceed to define collective operations over a set of sparse
vectors located at the nodes. We focus on allgather
and allreduce as defined by the MPI
specification~\cite{UsingAdvancedMPI}. We support arbitrary
coordinate-wise associative reduction operations for which a
neutral-element can be defined. (By \emph{neutral} we mean that the element
which does not change the result of the underlying operation, e.g., $0$ for the sum operation.)

\paragraph{Analytical Model.}
\label{subsec:Assumptions}
We assume bidirectional, direct point-to-point communication between the
nodes, and consider the classic Latency-Bandwidth ($\alpha$-$\beta$) cost model: the cost of
sending a message of size $L$ is $T(L) = \alpha + \beta L$,
where both $\alpha$, the latency of a message transmission, and
$\beta$, the transfer time per word, are constant. $L$ represents the
message size in words. When sending sparse items, let 
$\beta_s$ be the transfer time per sparse index-value pair and $\beta_d
< \beta_s$ the time per word.

Given this setting, the goal is to perform a collective operation over
the elements present initially at every node. That is, each node should
obtain the correct result locally, i.e., the element-wise sum over the
$N$ dimensions in the allreduce case, while minimizing the total
communication costs, measured in the $\alpha$-$\beta$ model.

\paragraph{Assumptions.}
For simplicity, we will assume that each node initially has $k$ elements: $\forall i:\,\vert
    H_i \vert = k$; $P$ is a power of $2$, $P > 4$; and $N$ is divisible by $P$.
We discuss these assumptions and relax
them in the supplementary material. 

\subsection{Communication Algorithms}

\newcommand{\ssarone}{SSAR\_Recursive\_double}
\newcommand{\ssartwo}{SSAR\_Split\_allgather}
\newcommand{\dsar}{DSAR\_Split\_allgather}

We proceed with the composition of sparse streams into sparse collective
communication algorithms. For this, we modify two existing dense
algorithms to efficiently work with sparse streams. 
In an allreduce operation, each node $i$ has a vector $x_i \in
\mathbb{R}^N$ and the operation computes the element-wise sum
$x=\sum_{i=0}^N x_i$ of all distributed vectors such that a copy of $x$
is available at each node after the operation.  
%
%

Allreduce can be implemented in many ways, for example, the nodes could
collaborate to compute the result at a single node (reduce) followed by
a broadcast or each node sends its $x_i$ to all nodes (allgather). 
If the vector $x_i$ is dense (the traditional case), the algorithms are
well understood and implemented in MPI
libraries~\cite{hoefler-moor-collectives}. 
However, we now explore the case where $x_i$ is sparse such that not all
elements need to be sent. We do not assume global information, i.e.,
none of our algorithms requires knowledge about the amount of data
contributed by each node, nor about the distribution of non-zero
indices.  

Yet, we require the user to have some rough idea about $\mathcal{K}$,
the final size of the result. This is often easily observable and we
will differentiate two types of instances: In \emph{static sparse
allreduce} (SSAR), $\mathcal{K}$ remains below $\delta$, such that we
will never switch to a dense representation. 
Conversely, in \emph{dynamic sparse allreduce} (DSAR) instances, where
$\mathcal{K} \geq \delta$, we will start with a sparse and switch to a
dense representation at some point during the collective operation.



If we assume that the number of non-zero indices is identical on all
nodes, we can distinguish two extreme cases: (1) none of the zero
elements overlap, i.e., $H_i \cap H_j = \emptyset\,\forall i,j$ and (2)
all elements overlap fully, i.e., $H_i = H_j\,\forall i,j$. The first
case is the case of maximum fill-in, at the end, $x$ will have $kP$
non-zero elements. If it would be known that no elements overlap, then
the sparse allreduce could be implemented efficienlty with a simple
allgather operation because no computation is necessary. Similarly, the
second case is equivalent to a dense allreduce of size $k$. Any other
possible distribution of non-zero indices lies in between these two
extremes. 
We can now bound the communication time from above in case (1) as
$\log_2(P)\alpha + (P-1)k\beta_d$~\cite{chan2007collective} 
and from below in case (2) as 
$\log_2(P)\alpha + 2\frac{P-1}{P}k\beta_d$~\cite{chan2007collective}. We
note that the latter communication bound is only valid for negligible
computational cost. 

\begin{lemma}
The time $T$ for sparse allreduce is bounded by $T\geq \log_2(P)\alpha +
(P-1)k\beta_d$ if $\mathcal{K} = k P$, and $T\geq \log_2(P)\alpha +
2\frac{P-1}{P}k\beta_d$ assuming that $\mathcal{K} = k$ and
computation for reduction is perfectly parallelized.
\end{lemma}


In practice, allreduce implementations switch between different
implementations depending on the message size and the number of
processes~\cite{thakur2003improving}. We distinguish between two cases:
small messages and large messages on a moderate number of processes. 

\subsubsection{The Small Data Case} 

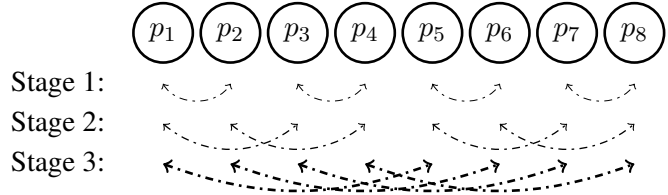
\begin{figure}[htbp]
	\centering
	\begin{tikzpicture}[node distance=0.07cm, line width=1pt]
	\tikzstyle{process} = [draw, circle, minimum width = 0.3cm, minimum height = 0.3cm];
	
	\node[process] (p10) {$p_1$};
	\node[process, right = of p10] (p20) {$p_2$};
	\node[process, right = of p20] (p30) {$p_3$};
	\node[process, right = of p30] (p40) {$p_4$};
	\node[process, right = of p40] (p50) {$p_5$};
	\node[process, right = of p50] (p60) {$p_6$};
	\node[process, right = of p60] (p70) {$p_7$};
	\node[process, right = of p70] (p80) {$p_8$};
	
	\def \n {8}
	
	\def \delta {0.0 cm}
	\foreach \s in {1,...,\n}
	{
		\node[below = \delta of p\s0] (p\s1) {\phantom{$p_\s$}};
	}
	
	\def \ldelt {0.3cm}
	\node[left = \ldelt of p11] (S1) {Stage 1:};
	
	\def \delta {0.0 cm}
	\foreach \s in {1,...,\n}
	{
		\node[below = \delta of p\s1] (p\s2) {\phantom{$p_\s$}};
	}
	
	\node[left = \ldelt of p12] (S2) {Stage 2:};
	
	\def \delta {0.0 cm}
	\foreach \s in {1,...,\n}
	{
		\node[below = \delta of p\s2] (p\s3) {\phantom{$p_\s$}};
	}
	
	\node[left = \ldelt of p13] (S3) {Stage 3:};
	
	\draw[<->,dash dot, line width=0.25pt] (p11.center) to [out=300,in=240] (p21.center);
	\draw[<->,dash dot, line width=0.25pt] (p31.center) to [out=300,in=240] (p41.center);
	\draw[<->,dash dot, line width=0.25pt] (p51.center) to [out=300,in=240] (p61.center);
	\draw[<->,dash dot, line width=0.25pt] (p71.center) to [out=300,in=240] (p81.center);
	
	\draw[<->,dash dot, line width=0.5pt] (p12.center) to [out=320,in=220] (p32.center);
	\draw[<->,dash dot, line width=0.5pt] (p22.center) to [out=320,in=220] (p42.center);
	\draw[<->,dash dot, line width=0.5pt] (p52.center) to [out=320,in=220] (p72.center);
	\draw[<->,dash dot, line width=0.5pt] (p62.center) to [out=320,in=220] (p82.center);
	
	\draw[<->,dash dot] (p13.center) to [out=340,in=200] (p53.center);
	\draw[<->,dash dot] (p23.center) to [out=340,in=200] (p63.center);
	\draw[<->,dash dot] (p33.center) to [out=340,in=200] (p73.center);
	\draw[<->,dash dot] (p43.center) to [out=340,in=200] (p83.center);
	
	\end{tikzpicture}
	\vspace{-0.25cm}
	\caption{Static Sparse allreduce: Recursive doubling - Increasing amount of sparse data in every stage}
	\label{fig:ssar_rec_dbl}
	\vspace{-0.1cm}
\end{figure}

When the overall reduced data is small, latency dominates the bandwidth
term. 
In this case, we adopt a \emph{recursive doubling} technique: in the
first round, nodes that are a distance $1$ apart exchange their data and
perform a local sparse stream reduction. 
In the second round, nodes that are a distance $2$ apart exchange their
reduced data.  
Following this pattern, in the $t$-th round, nodes that are a distance
$2^{t-1}$ apart exchange all the previously reduced $2^{t-1}k$ data
items. 
This behavior is illustrated in Figure~\ref{fig:ssar_rec_dbl}. 
The recursive doubling technique can also be used for solving
\emph{dense} allreduce and allgather problems~\cite{hoefler-moor-collectives}. 

The resulting latency for the \ssarone{} algorithm is $L_1(P) =
\log_2(P)\alpha,$ as there are $\log_2(P)$ stages. This is
latency-optimal and data-independent. The bandwidth term varies with the
sparsity pattern between the extremes discussed above: 
$$L_1(P) + \log_2(P)k\beta_s \leq T_{ssar\_rec\_dbl} \leq L_1(P) +
(P-1)k\beta_s.$$
The lower bound is reached when the $k$ indices fully overlap.
Therefore, at every stage, $k$ items need to be transmitted as the
intermediate results maintain constant size.  The upper bound is given
when the indices do not overlap at all.  Therefore, at stage $t$, the
number of items transmitted is $2^{t-1}k$. Taking the sum, we get
$$\sum_{i=1}^{\log_2(P)}2^{i-1}k = k\frac{2^{\log_2(P)} - 1}{2-1} =
k(P-1).$$

\subsubsection{The Large Data Case} 

When the data is large, dense allreduce implementations make use of
Rabenseifner's algorithm~\cite{rabenseifner2004optimization}, which has
two steps: The first is a Reduce-Scatter step, that partitions the
result vector across nodes, assigning a partition to each node. 
This is implemented by a \emph{recursive halving}
technique~\cite{rabenseifner2004optimization}. 
In the second step, the reduced answers are gathered to all other nodes
by calling a recursive doubling algorithm as described above. 
This two step algorithm has a total runtime of $$T_{ar\_rab} =
2\log_2(P)\alpha + 2\frac{(P-1)}{P}k\beta_s,$$ 
reaching the lower bound on the bandwidth term and off by a factor
$2$ on the latency term.

Our sparse allreduce for large data is inspired by this dense algorithm. 
It split the execution into two steps: a \emph{split} phase and a
\emph{sparse allgather} phase. 
In the \emph{split} phase, we uniformly split the space dimension $N$
into $P$ partitions and assign to each node the indices contained in the
corresponding partition. 
We split each sparse vector at its node and directly send each subrange
of indices to the corresponding recipient in a \emph{sparse} format. 
This direct communication comes at a higher latency cost, which we
mitigate by using non-blocking send and receive calls. 
%
%
Each node then reduces the data it received and builds the result for
its partition. 
In the second phase, the data has to be gathered to all other nodes with
a simple (concatenating) sparse allgather.

%
%
%
%
%

Obtaining runtime bounds for \ssartwo{} is slightly more involved. The
\emph{split} part takes time $$(P-1)\alpha + 0\beta_s \leq T_{split}
\leq (P-1)\alpha + k\beta_s.$$ 
Notice that both extremes imply that each node has $k$ items for the
sparse allgather and thus $\mathcal{K} = k P$.
For this second step in the algorithm to be optimal, every node must
have an intermediate result of size $\frac{k}{P}$, as we want the final
result to have a size $\mathcal{K} = k$ and the communication to be
equally distributed. 

For every node to have an intermediate result of the desired size, we
know that each node has to send at least $\frac{P-1}{P}k$ items to other
nodes. Otherwise, if every node has exactly $k$ items, we reach the
upper bound for the result size of $\mathcal{K} = k \times P$. So we get
$$L_1(P) + \frac{P-1}{P}k\beta_s \leq T_{sparse\_ag} \leq L_1(P) +
(P-1)k\beta_s.$$
The algorithm latency is again data-independent: $$L_2(P) = (P-1)\alpha
+ L_1(P).$$
Combining these terms yields 
$$L_2(P) + 2\frac{P-1}{P}k\beta_s \leq T_{ssar\_split\_ag} \leq L_2(P) + Pk\beta_s.$$

\subsubsection{The Dynamic Case: Switching to Dense}

\paragraph{Analysis.} The discussion so far focused on the case where maintaining a sparse
representation is efficient. 
However, as we gather data, the size of the result $\mathcal{K}$ might
become larger than the sparsity-efficient threshold $\delta$, in which
case we switch to a dense representation. 
We call this the \emph{dynamic} version of the problem (DSAR).  
The first result regarding this case is negative: the bandwidth savings due to sparsity are 
 limited to a constant improvement relative to the dense case. 

Assume the final size of the reduction result $\mathcal{K}$ is
larger than the threshold $\delta$, where a sparse representation is efficient. Let $\mathcal {K} \geq \delta = \kappa N$ be the size of the final reduction, which is too large to allow for a sparse representation (e.g. $\delta \log N \geq 1$). The algorithm will therefore switch to a dense representation at some point during the reduction
operation. Additionally, we want to avoid unnecessary computation and, 
following~\cite{chan2007collective}, we assume equally distributed optimal 
computation among the nodes during the reduction process. We further know that 
every node has to send $k$ elements to at least one other node and receive at least the other $\delta-k$ items of the dense final result. 
Following Chan et al.~\cite{chan2007collective} we can prove the following claim:
\vspace{-0.5em}
\begin{lemma}
	Any algorithm solving the DSAR problem needs at least time $\log_2(P)\alpha + \delta\beta_d$, where the lower bound on the bandwidth required is at least $\frac{1}{2}\kappa$ that of any bandwidth-optimal fully dense allreduce algorithm, with $\kappa = \frac{\delta}{N}$.
	\label{lem:dsar_min_dur}
\end{lemma}
\vspace{-0.5em}
\begin{proof}
	The optimal latency term is identical to the fully dense allreduce lower bound
	given by Chan et al.~\cite{chan2007collective}. The fully dense allreduce with 
	$k=N$ has a lower bound of $2\frac{P-1}{P}N\beta_d$ on the bandwidth, if computation is equally distributed. Based on the previous assumptions, the DSAR problem has a minimum bandwidth term of $\delta\beta_d$, which yields the $\frac{1}{2}\kappa$ factor as a lower bound.
\end{proof}
\vspace{-0.5em}

\paragraph{Algorithm.} Based on these insights, our solution for DSAR adapts the previous
two-stage algorithm to exploit the fact that every reduced split will
become dense. \dsar{} hence receives the data in a sparse format from
all the other nodes in the first phase, then switches the representation
and performs a dense allgather in the second stage. Here, we can
leverage existing implementations, which are highly optimized to perform
this second step with dense data. Based on the known times needed by
those algorithms, which are obviously independent of the input density,
we derive the running time for our algorithm given both extremes. The
latency is again $L_2(P)$. Combined, we get $$L_2(P) +
\frac{P-1}{P}N\beta_d \leq T_{dsar\_split\_ag}$$and
$$T_{dsar\_split\_ag} \leq L_2(P) + k\beta_s + \frac{P-1}{P}N\beta_d.$$

Another interesting observation following Lemma~\ref{lem:dsar_min_dur} is the fact, 
that by exploiting sparsity alone, and if the end-result is not efficiently storable
in a sparse format compared to the dense representation, the achievable speedup of 
a sparse allreduce is at most $\frac{2}{\kappa}\times$ (with $\kappa = 0.5$, this yield
a max speedup of $4\times$) compared to a fully dense algorithm. Other representation 
reduction techniques are needed in order to achieve higher speedups.


\vspace{-1em}
\section{Supporting Low-Precision Communication}

In the previous sections, we showed that the bandwidth cost in the dynamic case is lower bounded by a constant fraction of the bandwidth cost of a dense reduction, and we provided experimental evidence in the corresponding section that this case is in fact likely in large-scale deployments. 

Therefore, to further reduce bandwidth cost, \mml{} supports lower-precision data
representation for the outputs  (2, 4, and 8 bits per entry), using stochastic 
quantization as defined in the QSGD scheme~\cite{alistarh2016qsgd}. This quantization scheme provably preserves the convergence of the SGD  algorithm. Due to space constraints, we only present an outline of the scheme and its implementation.  

In brief, to implement QSGD quantization, each (dense) stream is
split into \emph{buckets} of size $B$ (in the order of 1024 consecutive entries) and each
bucket is quantized independently and  stochastically to the given number of quantization levels. 
Thus, each bucket
corresponds to $B$ \emph{low-precision data items}, e.g., 4-bit integers,
packed to reduce space and a full-precision \emph{scaling factor}, which is used to provide a scale to all the entries in the bucket.  
We focus on low-precision to reduce the bandwidth cost of the \emph{dense}
case. In practice, we employ the low-precision data representation only in the second part of the  \dsar{} algorithm, where the data becomes dense. This allows us to reduce the bandwidth cost of this last step by a constant corresponding to the quantization.  

\vspace{-1em}
\section{Artifact and Additional Features}
\label{sec:additional-features}

\paragraph{Interface and Code.} The \mml{} library provides a similar interface to that of standard MPI calls, with the
caveat that the data representation is assumed to be a sparse stream.
Given this, the changes needed to port MPI-enabled code to exploit
sparsity through \mml{} are minor.  The library implementation consists
of around 2,000 lines of native \verb!C++11! (This does not include
infrastructure such as benchmarks and tests which raises the line
count by an order of magnitude). Adding \mml{} to CNTK required changing
around $100$ lines of code. 

\paragraph{Non-Blocking Operations.}
We also implement the previous algorithms in a \emph{nonblocking
way}, similar as specified for nonblocking collectives in MPI-3~\cite{libnbc, hoefler-sc07}.  Specifically, we allow a thread to
trigger a collective operation, such as allreduce, in a nonblocking way.
This enables the thread to proceed with local computations while the
operation is performed in the background~\cite{hoefler-ib-threads}. 
%
%
%
%

\begin{figure*}[htbp]
	\centering
	\includegraphics[width=0.355\textwidth,height=\textheight,keepaspectratio]{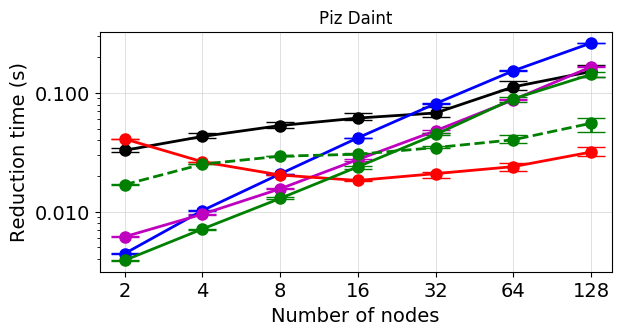}
	\includegraphics[width=0.575\textwidth,height=\textheight,keepaspectratio]{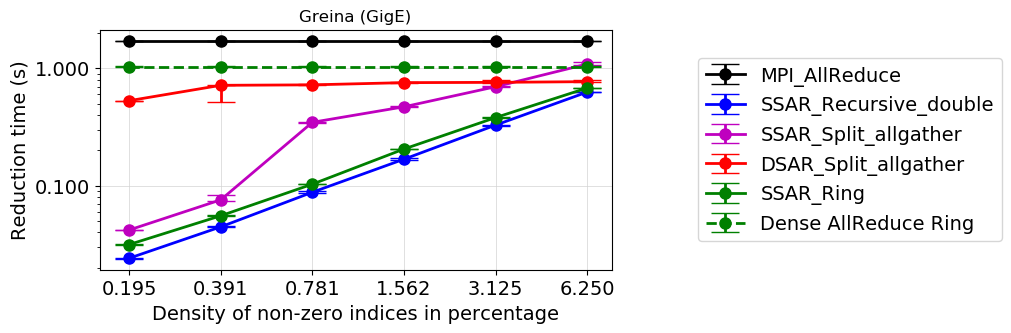}
	\vspace{-0.4cm}
	\caption{Reduction time versus number of nodes on Daint ($N=16\textrm{M}$ and $d=0.781\%$), and 
		reduction time versus data density on Greina ($N=16\textrm{M}$ and $P=8$) for various algorithms.}
	\label{fig:ComparePipeline}
	\vspace{-0.25cm}
\end{figure*}

%
%

\paragraph{MPI-OPT.} MPI-OPT is a framework we developed from scratch to run 
distributed optimization algorithms such as SGD. It is written in native \verb!C++11!, and can link external libraries such as \mml{} and MPI for communication. 
MPI-OPT implements parallel stochastic optimization algorithms, like gradient and coordinate
descent, on multiple compute nodes communicating via any MPI library,
with low overhead. It implements efficient distributed partitioning of any dataset converted in the predefined format using MPI-IO, 
data-parallel optimization on multiple compute nodes, with efficient multi-threading inside each node, parametrized learning rate adaptation strategies, 
as well as customizations to use \mml{} as the communication layer between nodes allowing for sparse, dense, synchronous, and asynchronous aggregation.

\paragraph{The Microsoft Cognitive Toolkit (CNTK).} For large-scale
neural network training, we modify  
CNTK~\cite{CNTK} v2.0 to use \mml{} as its communication layer. CNTK is a computational platform optimized for deep learning. 
The general principle behind CNTK is that neural network operations  are described by a directed computation graph, in which leaf nodes represent input values or network parameters, and internal nodes represent matrix operations on their children. 
CNTK supports and implements most popular neural network architectures. To train such networks, CNTK implements stochastic gradient descent (SGD) with automatic differentiation. 
The CNTK baseline supports parallelization across multiple GPUs and servers, with efficient MPI-based communication.

\vspace{-1em}
\section{Experiments}


\paragraph{Setup.} 
We now validate \mml{} on real world applications and synthetic
experiments.
Complete code and experimental logs will be made available once the
paper is de-anonymized.
Our experiments target two scenarios:
supercomputing and cloud computing.  For the first setting, we execute
on the CSCS Piz Daint supercomputer~\cite{PizDaint}, with Cray XC50
nodes, each of which has a 12 cores HT-enabled Intel Xeon E5-2690 v3 CPU
with 4GB RAM and an NVIDIA Tesla P100 16GB GPU. Piz Daint is currently
the most powerful supercomputer in Europe and has a
high-performance Cray Aries interconnect with a Dragonfly network topology.
We use multiple nodes using relatively older NVIDIA K80 GPUs connected
through Gigabit Ethernet to simulate a standard cloud deployment, but ensuring no background traffic. We perform additional tests on a distinct
cluster called Greina, with CX50 nodes and an InfiniBand FDR or 
Gigabit Ethernet interconnect and on a production-grade GPU cluster, described in the corresponding section.  


In all our experiments, the baseline will be the MPI allreduce implementation on the fully dense vectors. 
In general we make use of the default Open MPI installation. 
On Piz Daint, we compare against the custom Cray-MPICH installation, highly optimized by Cray. Since our problems usually have dimension $N > 65 \textrm{K}$,  
we fix the datatype for storing an index to an \verb!unsigned int!.

\vspace{-1em}
\subsection{Micro-Benchmarks}

We begin by validating our theoretical analysis on synthetic data, on the Piz Daint and Greina (GigE) clusters. 
We vary the data dimension $N$ and the data density $d$ as well as the number of nodes $P$. 
Based on the defined density, $k$ indices out of $N$ are selected uniformly at random at each node and are assigned a random value. 
We run our sparse allreduce algorithms in order to validate both correctness and the relative ordering of the derived analytical bounds. 
The choice of parameters is realistic (we pick data dimensions corresponding to common layer sizes in neural networks).  

For readability, graphs are in a log-log scale. 
As execution times are non-deterministic, we conduct five experiments with newly generated data, while running each one for ten times. 
Based on those 50 resulting runtime values, we state the 25 and 75 percentage quantiles. 
Results showing reduction times versus node count and density are given in Figure~\ref{fig:ComparePipeline}.  

Following the theoretical analysis, we expect the variant \ssarone{} to
perform best for a small amount of data, when latency dominates over the
bandwidth term. At higher node count $P$, data becomes larger, which
leads to less improvement of the algorithm \ssarone{} at the same number
of non-zero entries over the other variants.
Furthermore, the algorithm  \ssartwo{} dominates over the \dsar{} variant as long as the number of non-zero indices is relatively low compared to the overall reduced size. Both facts are visible in Figure~\ref{fig:ComparePipeline}, where we notice that the difference between \ssarone{} and \ssartwo{} is larger, when increasing the number of nodes compared to increasing density.
To show the impact of the network on performance, we did run identical tests on both Piz Daint and Greina (GigE) as in Figure~\ref{fig:ComparePipeline} on the right. The relative ordering remains comparable requiring less overall reduction time on high performance networks. 

Additionally, the experiments in Figure~\ref{fig:ComparePipeline} also compare our approaches against a ring-based MPI dense allreduce and its sparse counterpart. We note that, on a fast network and relatively small number of nodes, the ring-based algorithm is faster then any all other algorithms, but does not give any speedup at high number of nodes even at low density. As expected, \dsar{} offers improvement even at a relative large number of nodes, but only up to a constant factor. 



\begin{table}[hb]
	\begin{footnotesize}
		\centering
		\begin{tabular}{ @{} l | r | r | r @{} }
			\toprule
			Name & \# Classes & \# of samples & Dimension \\
			\midrule 
			URL~\cite{ma2009identifying} & 2 & 2 396 130 & 3 231 961 \\
			Webspam~\cite{webb2006introducing} & 2 &  350 000 & 16 609 143 \\
			\midrule 
			CIFAR-10~\cite{krizhevsky2009learning} & 10 & 60 000 & 32x32x3 \\
			\midrule
						ImageNet-1K~\cite{russakovsky2015imagenet} & 1000 & 1.3M & 224x224x3 \\
			\midrule
			ATIS~\cite{hemphill1990atis} & 128 & 4 978 \textit{s} / 56 590 \textit{w} & - \\
			\midrule
			Hansards~\cite{hansards} & - & 948K \textit{s} / 15 657K \textit{w} & - \\
			\bottomrule
		\end{tabular}
		\caption{Real World Application Datasets. \textit{s} stands for sentences (or pairs) and \textit{w} for words.  }
		\label{tbl:DataSets}
		\vspace{-0.5cm}
	\end{footnotesize}
\end{table}

\begin{table*}[htbp]
	\begin{footnotesize}
		\centering
		\begin{tabular}{ @{} l | l | l | r | r | l | r | r @{} }
			\toprule
			System & Dataset & Model & \# of nodes & Baseline Time (s) & Algorithm & Algo. Time (s) & Speedup \\
			\midrule
			\multirow{ 3}{*}{Piz Daint} & \multirow{ 3}{*}{Webspam} & LR & 32 & 24.0 (21.6) & \multirow{ 3}{*}{\ssarone} & 6.8 (3.5) & \textbf{3.53 (6.17)} \\
			&  & SVM & 32 & 16.2 (14.2) & & 6.5 (4.4) & \textbf{2.49 (3.23)} \\
			\midrule
			\multirow{ 3}{*}{Piz Daint}& \multirow{ 3}{*}{URL} & LR & 32 & 26.4 (25.8) & \multirow{ 3}{*}{\ssarone} & 7.5 (7.0) & \textbf{3.52 (3.69)} \\
			&  & SVM & 32 & 19.8 (19.3) &  & 5.6 (5.3) & \textbf{3.54 (3.64)} \\
			\midrule
			\multirow{ 2}{*}{Piz Daint}& Webspam & LR & 8 & 46.7 (37.9) & \multirow{ 2}{*}{\ssartwo} & 25.6 (15.8) & \textbf{1.82 (2.40)} \\
			& URL & LR & 8 & 37.7 (35.3) & & 20.9 (15.0) & \textbf{ 1.80 (2.35)} \\
			\midrule
			\multirow{ 2}{*}{Greina (IB)} & Webspam & LR & 8 & 65.2 (46.7) & \multirow{ 2}{*}{\ssartwo} & 36.3 (19.0) & \textbf{1.80 (2.46)} \\
			& URL & LR & 8 & 81.4 (44.7) & & 61.1 (24.9) & \textbf{1.33 (1.80)} \\
			\midrule
			\multirow{ 2}{*}{Greina (GigE)} & Webspam & LR & 8 & 768.0 (759.5) & \multirow{ 2}{*}{\ssartwo} & 37.9 (29.5) & \textbf{20.26 (25.75)} \\
			& URL & LR & 8 & 1045.0 (1004.6) & & 80.26 (42.2) & \textbf{12.65 (23.81)} \\
			\bottomrule
		\end{tabular}
		\caption{Distributed optimization using MPI-OPT. The times are averages for a full dataset pass, with the communication part in brackets. Speedup versus dense MPI is shown end-to-end, with communication speedup in brackets.}
		\label{tbl:ResutlsMpiOpt}
		\vspace{-0.5cm}
	\end{footnotesize}
\end{table*}

\vspace{-1em}
\subsection{Large-Scale Classification}
We use MPI-OPT to train linear classifiers (Logistic Regression, SVM) on large-scale classification datasets using SGD and stochastic coordinate descent (SCD).
The goal is to examine the runtime improvements by just exploiting the sparsity inherently present in the datasets and algorithms. More precisely, in these experiments, we do \emph{not} sparsify or quantize the gradient updates, but exploit the fact that data and hence gradients tend to be sparse for these tasks. 
The datasets are specified in Table~\ref{tbl:DataSets}. We examine the standard URL and Webspam high-dimensional binary classification datasets.

For SGD, the
samples have high sparsity since the features are
trigrams: while many such combinations exist, an item, e.g., a sentence,
can only have a very limited set of them present. This is extremely
common in text-based datasets. 
Since we are executing a task having a linear dependency between model and feature vector, this implies that the \emph{gradients themselves} will be sparse. 
Since communication is lossless,
convergence is preserved and we only report speedup of the communication
and overall training time. We run SGD with large batches ($1,000 \times
P$) for various combinations.  The achieved speed of MPI-OPT with the
best sparse reduction algorithm is reported in
Table~\ref{tbl:ResutlsMpiOpt}. (Communication speedup is reported in brackets.)

Additionally, we run
MPI-OPT's SCD implementation, which follows the distributed random 
block coordinate descent algorithm
of~\cite{wright2015coordinate}. We focus on optimizing
directly the primal problem in order to showcase the 
usage of \mml{} on other algorithms, ignoring the fact that
more sophisticated algorithms solving the dual problem
might exist~\cite{jaggi2014communication}. We run the optimization on the logistic
regression loss function for the URL dataset distributed on 8 nodes of
Piz Daint to achieve identical convergence compared to SGD. Every node
contributes 100 coordinates after every iteration. As the values calculated
by each node lie in different slices of the entire model vector, we
compare the runtime of an sparse allgather from \mml{} to its dense
counterpart. MPI-OPT with a dense allgather has an average epoch time of 49 seconds, with 24 seconds dedicated to communication. The sparse allgather executes a dataset pass (epoch) in 26 seconds on average, with 4.5 seconds spent in the communication layer. This implies an overall speedup of factor $\bm{1.8}\times$, due to a $\bm{5.3}\times$ speedup in communication time.

\paragraph{Comparison with Apache Spark.} As an exercise, we also compare MPI-OPT with Apache Spark v1.6, which is officially supported by CSCS~\cite{CSCSSpark}.  Comparison is performed on the same datasets; Spark uses its own communication layer which does not exploit sparsity. 

On Piz Daint, using $8$ nodes, MPI-OPT with \mml{} reduces the time to convergence on the URL dataset by ${63\times}$.
This is largely due to the reduction in communication time, which we measure to be of ${185\times}$. Concretely, the average epoch time is reduced from 378 seconds, with 319 seconds spent for communication, to an average of 6 seconds per epoch, whereof 1.7 seconds represent the communication time.
Compared to Spark, MPI-OPT with the standard Cray-optimized \emph{dense} allreduce has a $31\times$ speedup to convergence, due to a $43\times$ speedup in communication time. An epoch is executed in 13 seconds on average, with 8.6 seconds spent on communication. We further investigated these speedups on an 8-node research cluster with a Gigabit Ethernet interconnect. 
Using MPI-OPT, the average
training time per epoch drops from 1,274 seconds (Spark) to 14 seconds (${86\times}$). On the communication part,
the time per epoch drops from 1,042 seconds to 6 seconds. The communication time and
overall speedup of a \emph{dense} allreduce over Spark's communication layer are both of factor ${12\times}$.

\paragraph{Discussion.} The Spark comparison should be taken with a grain of salt, since Spark implements additional non-trivial features, notably \emph{fault-tolerance}. However, we believe our results show conclusively that  sparsity support can provide significant savings in this large-scale classification scenario, where sparsity is naturally present. 

\begin{figure*}[htbp]
	\centering
	\subcaptionbox{Training Accuracy for ResNet-110 Model on CIFAR-10.\label{fig:train_cifar}}[0.45\textwidth]{
		\includegraphics[width=0.4\textwidth,height=\textheight,keepaspectratio]{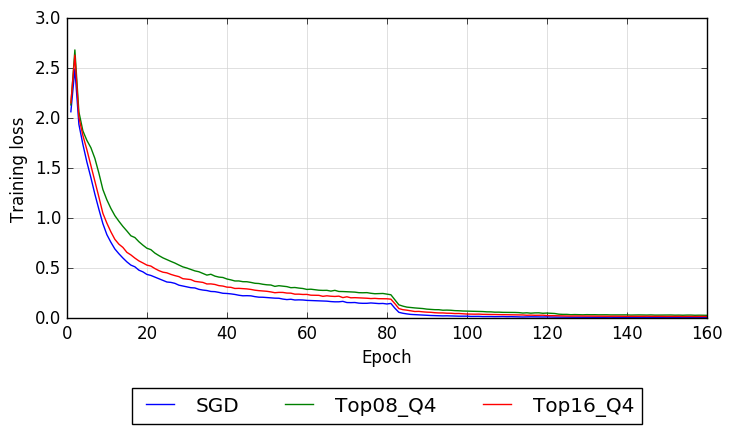}
	}
	\subcaptionbox{Training Accuracy LSTM Model on ATIS Dataset.\label{fig:train_atis}}[0.45\textwidth]{
		\includegraphics[width=0.4\textwidth,height=\textheight,keepaspectratio]{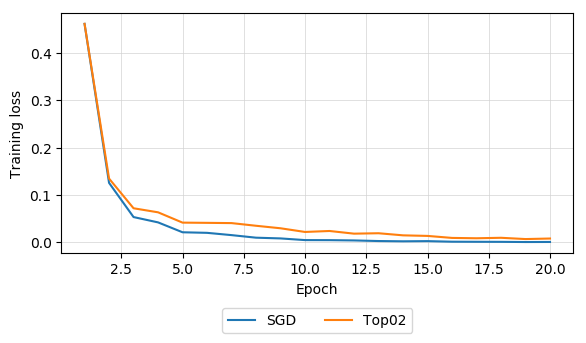}
	}
	\vspace{-0.3cm}
	\caption{Train Accuracy for Sparsified (and Quantized) Versions Vs. Full Dense SGD.}
	\label{fig:train_accuracy}
\end{figure*}

\begin{figure*}[htbp]
	\centering
	\subcaptionbox{Top5 Training Error.\label{fig:train_4xresnet}}[0.45\textwidth]{
		\includegraphics[width=0.37\textwidth,height=\textheight,keepaspectratio]{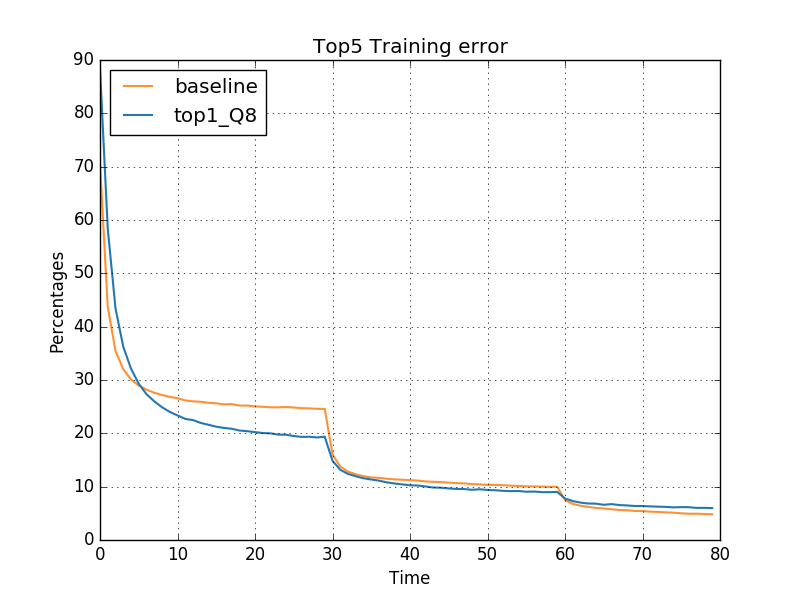}
	}
	\subcaptionbox{Top5 Validation Error.\label{fig:test_4xresnet}}[0.45\textwidth]{
		\includegraphics[width=0.37\textwidth,height=\textheight,keepaspectratio]{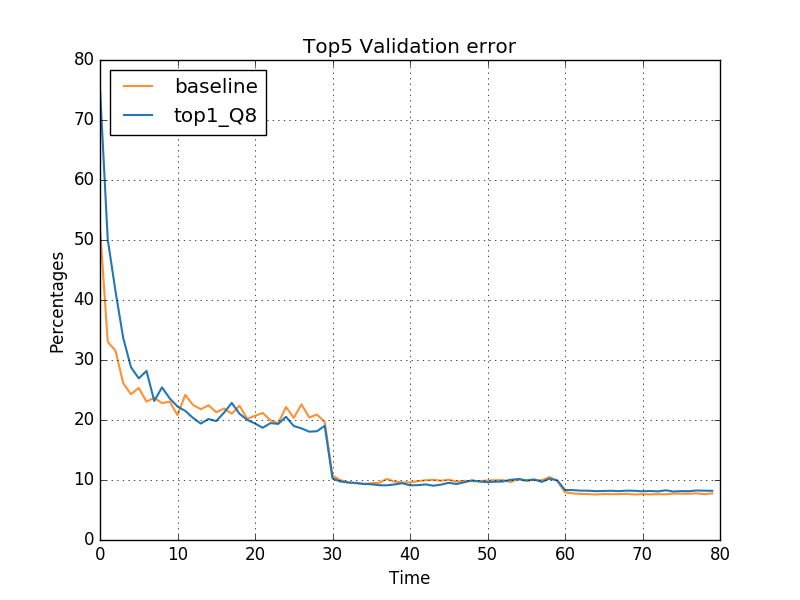}
	}
	\vspace{-0.3cm}
	\caption{Train and Validation Error for $4\times$ Wide ResNet for the baseline (orange) versus TopK Quantized SGD implemented in \mml{}.}
	\label{fig:wide-resnet}
\end{figure*}

\vspace{-1.1em}
\subsection{Training Deep Neural Networks}

In this section, we examine the applicability of \mml{} for distributed training of deep neural networks in CNTK, on academic datasets. (We present results on a larger tasks in the next section.)  
To exploit sparsity, we implement the Top-$k$ SGD algorithm~\cite{dryden2016communication, sun2017meprop, aji2017sparse} with low-precision support. 
The resulting protocol is provided in Algorithm~\ref{algo:qtopk-sgd}.  
We execute three types of tasks: \emph{image classification} on the
CIFAR-10 dataset, \emph{natural language
understanding} on the ATIS corpus and \emph{machine translation} on the
Hansards dataset. (See 
Table~\ref{tbl:DataSets} for details.) 
For vision, we train the ResNet-110 architecture~\cite{he2016deep}. For natural language understanding and machine translation we use an encoder-decoder network consisting of two LSTM~\cite{hochreiter1997long} cells each. 
We use the default hyper-parameters for single-GPU 32-bit full accuracy convergence in all our experiments, as provided in the open-source CNTK 2.0 repository~\cite{CNTKEX}. For completeness, these parameters are provided in the Supplementary Material. ~ 
For CIFAR-10 we select $k=8$ and $16$ entries
from every bucket of $512$ consecutive elements ($\sim 3\%$ density), and stochastically quantize the values to 4-bit precision. For ATIS we select $k = 2$, and $k = 4$ for Hansards, entries out of each bucket of $512$ ($\sim 0.4\%$ and $\sim 0.8\%$ density), using no additional quantization strategy. 
Top-$k$ selection and quantization are implemented using optimized GPU kernels, and communication is done layer-wise using non-blocking calls; this ensures that the impact on overall computation is minimal ($<1\%$).

To illustrate the bandwidth reduction, we note that the LSTM model we use for ATIS has approximately 20M parameters, which total approximately 80 MB in full precision, which would need to be transmitted every upon every minibatch. By contrast, the compressed gradient received by every node in \mml{} totals less than 0.5 MB.

\paragraph{Accuracy \& Speedup. } 
The key metric we track is the \emph{accuracy} of the converged models. 
For this, we note that on image classification (CIFAR-10), the model is able to recover virtually the same accuracy, both in terms of training and test error versus the number of epochs. 
Specifically, the end accuracy matches that of the full-precision baseline when  selecting $k=16$ out of every $512$ elements, and for $k=8 / 512$ the accuracy is $1\%$ \emph{above} the 32-bit variant as visible in Figure~\ref{fig:train_cifar}. For both ATIS and Hansards tasks, training and test metrics (losses and BLEU scores) are within $1\%$ of the full-precision baselines, as shown for ATIS in Figure~\ref{fig:train_atis}.

Examining end-to-end training speedup, on the CIFAR-10 task we achieve an overall speedup of factor
${1.12\times}$ to full convergence with 8 nodes on Piz Daint versus the full-precision baseline. Training ATIS for 20 epochs, and Hansard for 5200 iterations (as standard), we are able to reduce the overall training time on Piz Daint by a factor $5.99\times$ for ATIS, and $1.5\times$ for Hansard respectively. The variance in these speedup numbers is explained by the varying ratios of communication and computation of the models: for the models we employ on CIFAR-10 and Hansards, computation dominates communication, whereas this ratio is inverted for ATIS, in which case reducing communication has a much larger impact on end-to-end training time.

\begin{figure*}[htbp]
	\centering
	\subcaptionbox{Accuracy versus training time numbers for $6$ training passes over the entire dataset, recording training error (CE loss). Validation results (word-error-rates) are discussed in the text.\label{fig:production-accuracy-vs-time}}[0.45\textwidth]{
		\includegraphics[width=0.4\textwidth,height=\textheight,keepaspectratio]{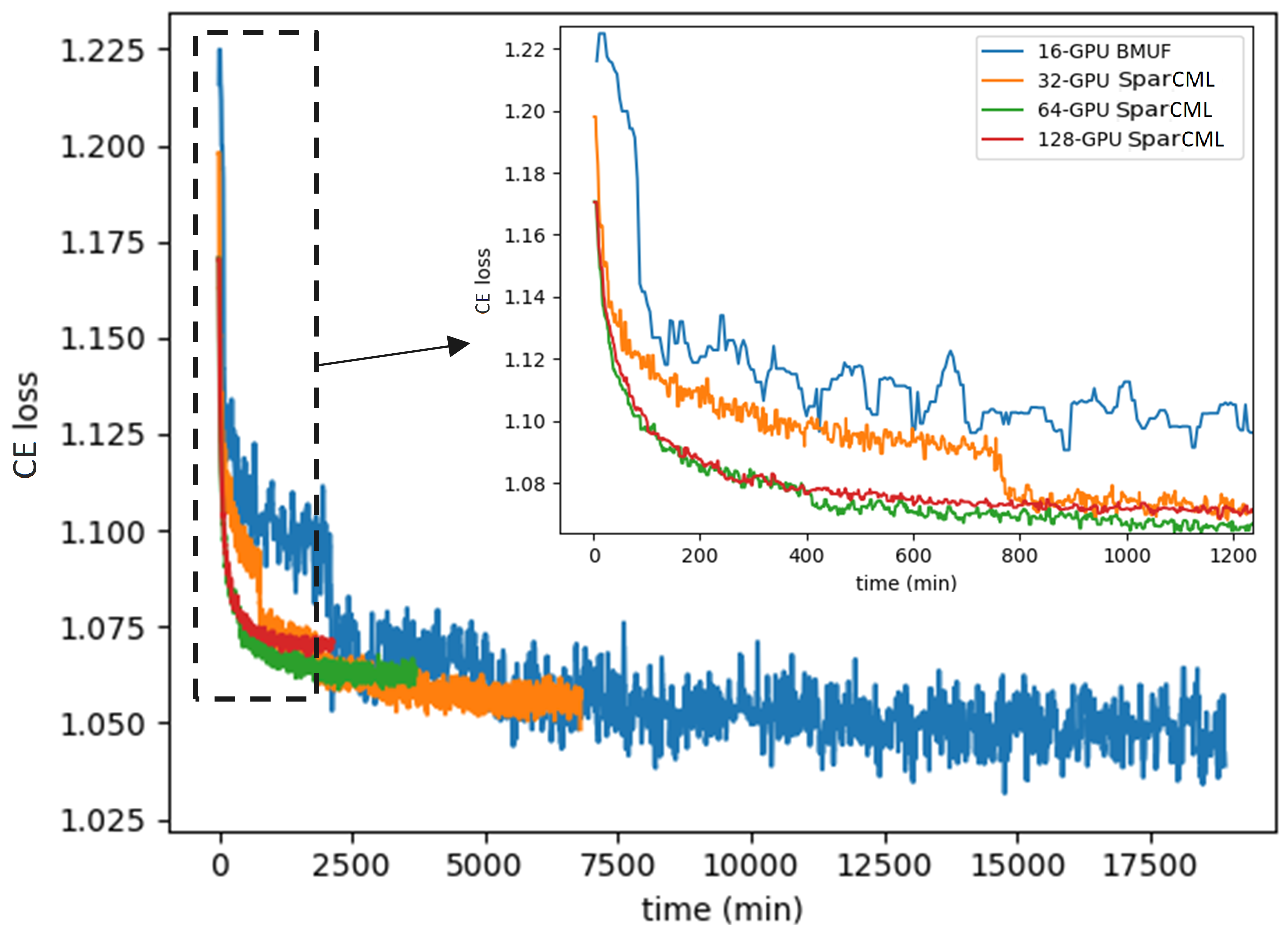}
	}
	\subcaptionbox{SparCML Scalability as a function of number of GPUs \label{fig:production-scalability}}[0.355\textwidth]{
		\includegraphics[width=0.305\textwidth,height=\textheight,keepaspectratio]{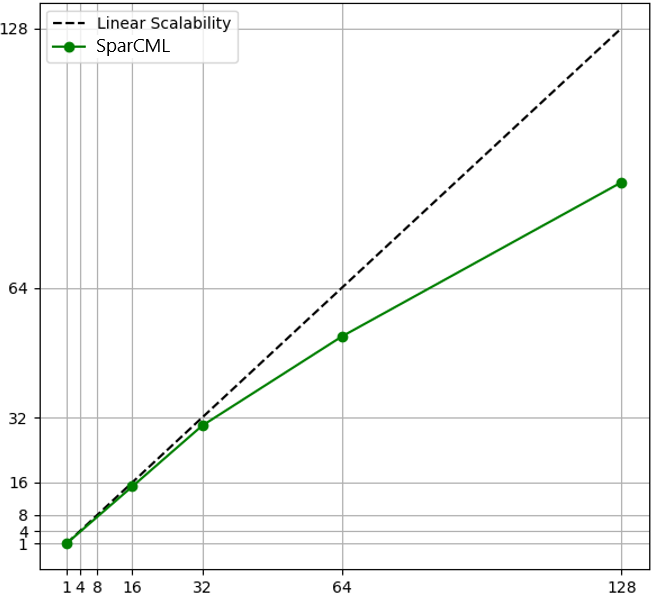}
	}
	\vspace{-0.3cm}
	\caption{Production Workload Speech Experiments.}
	\label{fig:ProductionWorkload}
\end{figure*}

\vspace{-1em}
\subsection{Large Workload Experiments}

\paragraph{ImageNet Experiments.} 
Our next experiment considers the applicability of our method in the context of large-scale image classification, in particular by training ResNet architectures on the standard ImageNet ILSVRC dataset, with 1000 target classes~\cite{russakovsky2015imagenet} (also known as ImageNet-1K). This experiment was executed on CSCS Piz Daint, using $64$ compute nodes, each with a P100 GPU. 

Our first target model is the classic ResNet50 architecture~\cite{he2016deep},  totalling approximately $25$ million parameters across $50$ layers. 
This model scales well when using the baseline Cray MPI implementation, as it benefits from the relatively low number of parameters, and from the fact that per-layer gradient transmission can be overlapped via non-blocking calls.  
For this model, our results in terms of scaling are negative: the runtime improvements due to layer sparsification to $99\%$ sparsity are of $\simeq 6\%$ ($1950$ seconds per epoch versus $2071$ for the Cray MPI baseline). Our profiling revealed that this negative result is due to several factors, in particular that: 
(1) For this parameter setting, gradients become dense during aggregation, which limits our speedup. We found that enforcing higher sparsity levels hurts model convergence, even if we implemented techniques such as momentum correction and warm-up training~\cite{lin2017deep} to alleviate this issue; 
(2) The overhead of sparsification and densification during TopK is non-negligible relative to the transmission cost of ResNet50 layers;
(3) Our implementation does not benefit from the additional parameter tuning of the proprietary Cray implementation. 
While items (2) and (3) can be addressed with additional implementation effort, item (1) strongly suggests that sparsification is not a universal solution for scaling to the very large node counts achieved by previous work for this type of CNN architecture~\cite{goyal2017accurate, you2017scaling}: since we cannot scale the sparsity linearly with the number of nodes without hurting model convergence, gradients naturally will become dense at high node count, which limits the benefits of our method in this scenario. 

Our second target model class for this task is \emph{wide} residual models~\cite{zagoruyko2016wide}. These models are variants of ResNet architectures, where  the only difference is that the number of \emph{channels} in each block is multiplied by a constant factor. It has been empirically found that shallow variants of wide models can achieve similar or better levels of accuracy as considerably deeper architectures~\cite{zagoruyko2016wide}, and that they are less sensitive to hyperparametrization, and in particular to large-batch training~\cite{chen2018effect}.  
Due to their increased capacity, wide residual networks are popular when transferring to more complex tasks, such as ImageNet-10K and 22K~\cite{codreanu2017scale}. 
In particular, we focus on training the 4xResNet18 and 4xResNet34 models
(which have $4x$ the channels of their regular variants) on ImageNet-1K
using TopK SGD, with $K = 1/512$, that is, on average only the top
$0.2\%$ of parameter values are transmitted by each node. Each P100 GPU
can only process four images in a batch, leading to a global batch size of $512$ images. 
We emphasize that we employ standard hyperparameter values for training these networks--besides theglobal batch size of $512$ which is higer due to parallelization; in particular, our learning rate schedules are identical to the single-GPU case, and we perform no adjustments for sparsity, such as warmup or momentum correction.  

Convergence results for 4xResNet18 are presented in Figure~\ref{fig:wide-resnet}, for both training and validation accuracy. 
We notice that the final accuracy of the models differs by less than $0.9\%$ in terms of top-1 accuracy, and less than $0.5\%$ in terms of top-5 accuracy. At the same time,~\mml{} provides a speedup of $\simeq 2\times$ versus the Cray MPI baseline. 
Upon examination, this speedup is due almost entirely to the reduced aggregation time on the gradients on the last fully-connected layer of the network, which totals more than $2$M parameters on this wide variant. The results are similar for 4xResNet34: the speedup is of approximately  $\simeq 1.85\times$ versus the Cray MPI baseline, with accuracy difference of $0.8\%$ in terms of top-1 accuracy, and less than $0.4\%$ in terms of top-5 accuracy versus the fully-dense baseline. 
We note the faster loss reduction of TopK in the earlier stages of training, whereas the improvement saturates and inverts at the end of training. 
In sum, we conclude that gradient sparsity can indeed provide non-trivial speedups for wide residual networks, at the cost of a relatively minor decrease in accuracy, with no additional hyperparameter tuning.

\paragraph{Automated Speech Recognition.} 
The final test of our framework is on a state-of-the-art acoustic model for automated speech recognition (ASR), powering a popular digital personal assistant. 
Due to anonymization constraints, some details of our setup are omitted. 
The model we train is a state-of-the-art LSTM network with attention. The model has more than 60 million parameters, 2.4 million of which reside in the attention layer. 
We employ Top-$k$ SGD for the training of the attention layer, starting from a pre-trained LSTM network. 
The dataset consists of approximately 30,000 hours (3.5 years) of annotated speech. 
Our cluster deployment consists of 32 server nodes, each with four NVIDIA V100 GPUs, totalling 128 GPUs. 
Servers have an InfiniBand interconnect, and aggregation inside each node is performed via NVIDIA NVLink with NCCL~\cite{NCCL}. 

The baseline we compare against is training on 4 nodes, 16 GPUs in total, without sparsity or quantization, but employing a carefully-tuned instance of block-momentum SGD (BMUF)~\cite{Block-Momentum-SGD}. 
\emph{Higher node counts for this full-precision variant led to negative scalability
and, in some cases, divergence}. 
We note that this baseline already performs non-trivial communication reduction, since it communicates updates less frequently between nodes with respect to standard minibatch SGD.
(Standard minibatch SGD is infeasible on our setup due to the large model size and node count.)

We execute six passes over the entire dataset and register the time to complete the experiment and the final accuracy. 
The 16 GPU BMUF baseline takes approximately 14 days to complete. This variant increases the batch size linearly with the number of nodes (weak scaling).
We compare against our version of Top-$k$ SGD with~\mml{}, in which gradients are split into groups of 512 consecutive coordinates, out of which we select the 4 largest ones, which we transmit from each group, saving the rest locally. We aim to leverage the fact that, in this production setting, most updates will occur in the parameters of the attention layer. When executing this variant, we tuned the initial learning rate, and the batch size; in particular, we keep a fixed global batch size of $512$ samples, which is the same as for sequential training (strong scaling).

 Figure~\ref{fig:production-accuracy-vs-time} presents the results in error-versus-time format, where error is measured by standard cross-entropy (CE) loss, using our implementation, for 32, 64, and 128 GPUs. 
We highlight the fact that the sparse implementation is able to reach similar accuracy to the full-precision baseline in a fraction of the time: at 32 nodes (128 GPUs), we are able to reduce training time to $<1.8$ days.  
Figure~\ref{fig:production-scalability} illustrates the good scalability of the method. 
To further test accuracy, we also performed testing in terms of word-error-rate (WER) for the converged models, on validation sets. 
We found that the models trained with SparCML incur error rates that are less than $1\%$ higher than full-precision (but unscalable) training and can sometimes \emph{improve} accuracy by up to $1\%$.  
This trade-off is very advantageous for this application scenario, as it enables much faster model iteration times.

\paragraph{Hyperparameter Tuning.}
One important question regards the need for additional hyperparameter tuning when using the Quantized TopK algorithm. We note that, although we enforced sparse gradients, we have recovered accuracy under standard hyperparameter values even under high sparsity levels, in most cases. 
There are two notable exceptions: ResNet50 training, where high sparsity combined with large batch sizes induced significant accuracy loss, and the ASR experiment, where we have maintained a small global batch size to preserve convergence. These results suggest a non-trivial interaction between sparse gradients, batch size, and convergence, which we aim to investigate further in future work.

\vspace{-0.5em}
\section{Related Work}

There has recently been a tremendous surge of interest in distributed machine learning~\cite{TF, CNTK, MXNET}; see Ben-Nun and Hoefler~\cite{distdl-preprint} for a survey. In the following we focus on closely related techniques. 

\paragraph{Reduced Communication Techniques.} 
Seide et al.~\cite{seide20141} was among the first to propose quantization to reduce the bandwidth and latency costs of training deep networks. 
More recently, Alistarh et al.~\cite{alistarh2016qsgd} introduced a theoretically-justified distributed SGD variant called Quantized SGD (QSGD), which allows the user to trade off compression and convergence rate. 
We implement QSGD as a default quantization method.  
Dryden et al.~\cite{dryden2016communication} and Aji and Heafield~\cite{aji2017sparse} considered an alternative approach to communication reduction for data-parallel SGD, \emph{sparsifying} the gradient updates by only applying the top-$k$ components, taken at at every node, in every iteration, for $k$ corresponding to $<1\%$ of the update size. Since then, other references~\cite{sun2017meprop, lin2017deep} explored this space, showing that extremely high gradient sparsity ($<0.1\%$) can be supported by convolutional and recurrent networks with preserved accuracy, although maintaining accuracy requires   hyperparameter tuning. 

Our paper complements this line of the work by 1) considering stochastic quantization and sparsification in conjunction, and proving that the resulting technique still provably converges and is practically useful; 2) providing highly efficient sparsity and quantization support, with consistent runtime gains in large-scale settings, both for supercomputing and cloud computing scenarios. 	

\paragraph{Lossless Methods.} Factorization is a lossless compression technique~\cite{chilimbi2014project, xing2015petuum} that is effective in deep neural networks with large fully-connected layers, but less applicable in networks with large convolutional layers, which are quite common~\cite{he2016deep, szegedy2017inception}. 
A second lossless method is executing \emph{extremely large batches}, thus hiding the cost of communication behind larger computation~\cite{goyal2017accurate, you2017scaling, akiba2017extremely, yamazaki2019yet}. 
The compression methods in \mml{} are orthogonal to this direction, as they aim to reduce bandwidth cost given a fixed batch size; as we have observed experimentally, sparsification can be applied with little additional tuning at a fixed batch size. 
However, when distributing training to large node counts, batches become large, and the aggregated gradients become dense, as we usually cannot scale sparsity up linearly with the node count. 
Thus, large-batch sparse-gradient hyperparameter tuning would become necessary in such cases, which we leave for future work. s
We note that \mml{} already implements several optimizations which are common in the large-batch setting, such as merging gradients for adjoining layers (``tensor fusion''), or non-blocking operations~\cite{yamazaki2019yet}.

\paragraph{Communication Frameworks.} Several frameworks have been proposed for reducing communication cost of distributed machine learning. 
One popular example is NVIDIA's NCCL framework~\cite{NCCL}, which significantly reduces communication cost when the nodes are NVIDIA GPUs and the proprietary NVLINK interconnect is available, which is not the case in multi-node settings, such as supercomputing. Further, NCCL currently only implements a very restricted set of reduction operations.
In addition, there is a non-trivial number of frameworks customized to specific application scenarios, such as the Livermore Big Artificial Neural Network Toolkit (LBANN)~\cite{van2015lbann} or S-Caffe~\cite{awan2017s}.
While very efficient in specific instances, these frameworks do not usually leverage reduced-communication techniques, or sparsity. 

\paragraph{Sparse Reduction.} Hofmann and R\"unger~\cite{hofmann2008mpi}
propose a simple and effective runlength encoding approach for sparse
reductions.  We significantly extend this approach in the current work,
including the observation that data might become dense during the
reduction process and that an efficient and flexible data representation
must be provided in this case. 
Tr\"aff~\cite{traff2010transparent} proposes a general approach for implementing sparsity in MPI by 
ignoring neutral elements in MPI reductions. Our sparse allreduce implementation could be seen as a special case of this general approach, where we precisely specify the reduction algorithms, and carefully analyze the performance bounds for small and large message scenarios. In addition, \mml{} makes several additional contributions which are specific to machine learning applications, such as efficient low-precision support and integration with machine learning frameworks. 


Kylix~\cite{zhao2014kylix} considers sparse many-to-many
reductions in the context of computation over large scale distributed
graph data on community clusters. However, Kylix assumes knowledge of
the data distribution and performs multiple passes over the reduction,
which make it not applicable to our scenario.  Dryden et
al.~\cite{dryden2016communication} implement a sparse variant of the
classical allreduce algorithm via a pairwise reduce-scatter followed by
a ring-based allgather. The amount of data is kept constant at every
stage of their algorithm by re-selecting the top $k$ values and
postponing the other received values. We note that this ability to
preserve a local residual is specific to Top-$k$ SGD and that our
framework is more general. In terms of performance, their implementation
will provide similar results to our \ssartwo{} algorithm.

\vspace{-1em}
\section{Conclusions and Further Work}

We have described and analyzed \mml{}, a high-performance communication
framework that allows the user to leverage sparse and low-precision
communication in the context of machine learning algorithms.  \mml{} integrates easily into
 existing computational frameworks and can provide order-of-magnitude
speedups in several real-world applications. 
%
In future work, we aim to further investigate other distributed machine learning applications which can benefit from sparsity, and to further investigate the interaction between sparsity and other parallelization approaches, such as large-batch training.
We believe that the simple but effective sparsity schemes we described can play a significant role in reducing communication cost in future machine learning systems.

\section*{Acknowledgments} 

This project has received funding from the European Research Council (ERC) under the European Union’s Horizon 2020 research and innovation programme (grant agreement No 805223), and from a Swiss National Supercomputing Centre Small Development Project (code d94).


\bibliographystyle{acm}
\bibliography{references}

\newpage
\appendix

\section{Relaxation of Assumption in Section~4}

Even though the three assumptions given in Section 4 simplify the formulas in the subsequent analysis of the algorithms, they do not oversimplify the problem. Ignoring assumption (1) and having $k = \max_i \vert H_i \vert$, one gets an upper bound on each algorithm. This upper bound only makes sense if we assume approximately an equal number of non-zero elements at every node. Otherwise, one could imagine to design more specific algorithms. If assumption (2) does not hold, one can add two additional steps in front and at the end of every algorithm to reduce the number of participating nodes to the nearest lower power of two. Although this might not be optimal (a dissemination approach ~\cite{hensgen1988two} might be favorable), the cost increases by some constant value and thus, we still get an idea about which algorithm to prefer. If assumption (3) does not hold, each node gets responsible of $\lfloor \frac{N}{P} \rfloor$ items apart of the last one, which is responsible of $N - (P-1)\lfloor \frac{N}{P} \rfloor$ items.

\section{Stochastic Density Analysis}

We realize, the difficulty of designing any efficient algorithms comes from the fact that we neither know in advance the exact number of items every node contributes, nor the size any intermediate, or the final result will have. This data has to be communicated across the network. Those result sizes are not only dependent on the amount of data contributed by each node, but also alters with different positions of the non-neutral indices. 
If one assumes an underlying probability distribution of the non-zero elements, one can define the expected total number of non-zero elements $\mathbb{E}[\mathcal{K}]$. We therefore make use of $N$ Bernoulli random variables $X_{j} = 1$, if index $j \in \cup_{i=1}^P H_i$, and $X_{j} = 0$ otherwise, for $1 \leq j \leq N$. The random variable $Y = \sum_{j = 1}^{N} X_j$ then represents the total number of non-zero entries after having performed the reduction. By using the linearity property of the expectation, we get: $$\mathbb{E}[\mathcal{K}] = \mathbb{E}[Y] = \sum_{j=1}^{N} P\left(j \in \cup_{i=1}^P H_i \right).$$ The probability of any index $j$ being an element of a distinct set $H_i$ is given by the underlying distribution. It is true for any distribution that: 
\begin{multline*}
P\left(j \in \cup_{i=1}^P H_i \right) = \sum_{i=1}^{P}P\left(j \in H_i\right) - \sum_{i < k}P\left(j \in \left(H_i \cap H_k\right)\right) + \\\sum_{i < k < l}P\left(j \in \left(H_i \cap H_k \cap H_l\right)\right)\,\dots\,+ (-1)^{P-1}P\left(j \in \cap_{i=1}^{P}H_i\right).
\end{multline*}We further know from Union Bound that $P\left(j \in \cup_{i=1}^P H_i \right) \leq \sum_{i=1}^{P}P\left(j \in H_i\right)$, which gives us a valuable upper bound on the expected number of non-zero elements $$\mathbb{E}[Y] \leq \sum_{j=1}^{N}\sum_{i=1}^{P}P\left(j \in H_i\right).$$ This bound is tight if $\forall i < j:\,H_i \cap H_j = \emptyset$, which is the special case where the problem reduces to an allgather.

\subsection{Uniform Distribution}
\begin{figure}[htbp]
	\centering
	\includegraphics[width=0.4\textwidth,height=\textheight,keepaspectratio]{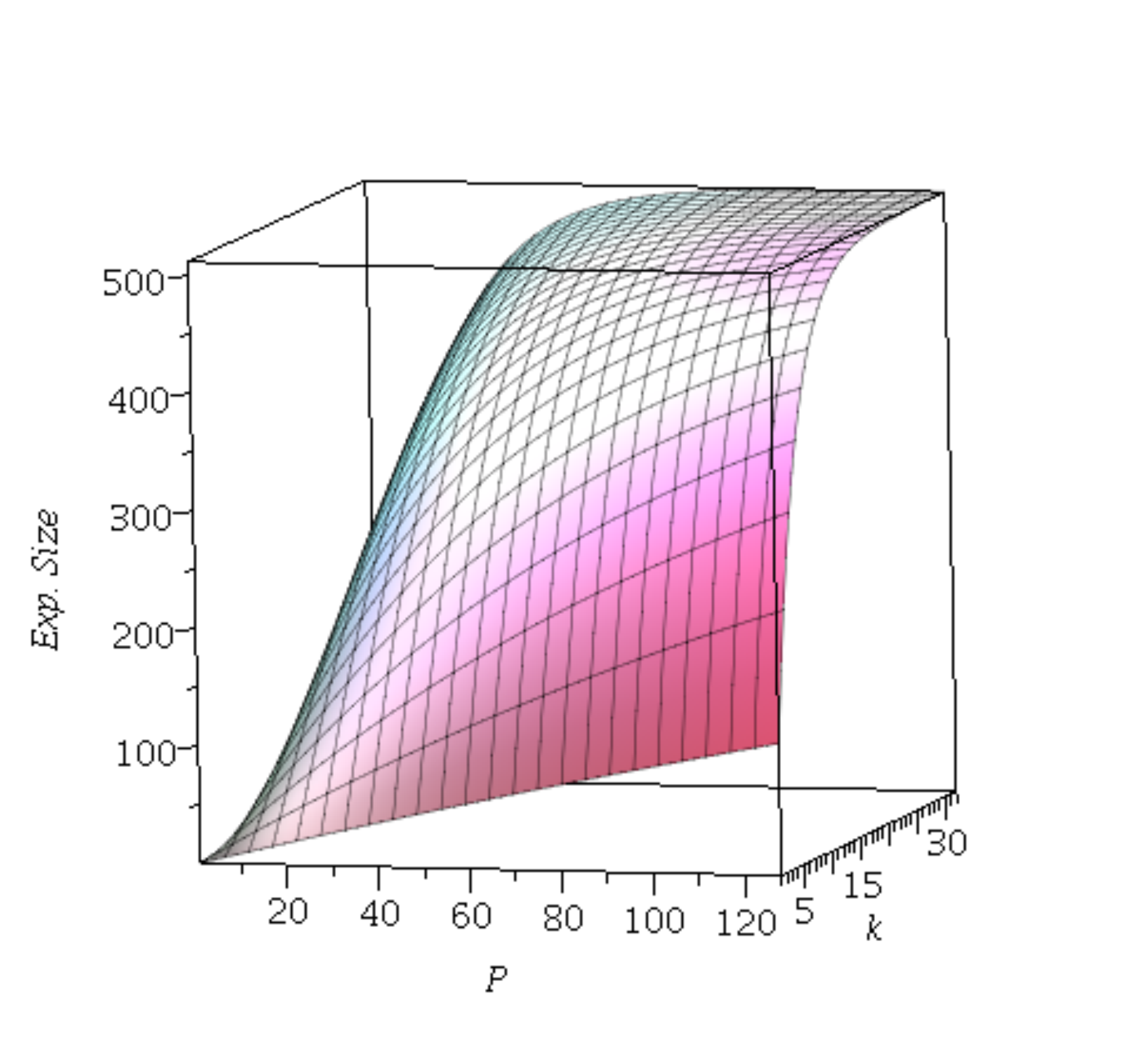}
	\caption{Expected size of reduced result assuming a uniform distribution and $N=512$}
	\label{fig:ExpSizeUnifDistrib}
\end{figure}

Having derived those formulas, we give more concrete values by assuming a uniform distribution. This use-case gives a worst-case scenario in terms of probabilistic growth of the intermediate results and it is reasonable to make this assumption, if every index is hit with probability higher than $0$.
For this, let $H_i$ consist of $k$ independent samples drawn from a uniform distribution $$j \sim \mathcal{U}\left(1,N\right) \quad \forall j \in H_i,$$ therefore $P\left(j \in H_i\right) = \frac{k}{N}$. This is independent of the two indices $i$ and $j$ in the above general formula, so $\mathbb{E}[\mathcal{K}] \leq N \times P \times \frac{k}{N} = P \times k$, which fits the non-probabilistic upper bound given earlier. For the uniform distribution one can give the exact expected number of elements by deriving a closed-form function utilizing the previous equations$$\mathbb{E}[\mathcal{K}] = f(k,N,P) = N \times \left(\sum_{i=1}^{P}(-1)^{i-1}{{P}\choose{i}}\left(\frac{k}{N}\right)^i\right).$$ Figure~\ref{fig:ExpSizeUnifDistrib} illustrates the multiplicative growth dependent on both inputs, the number of nodes $P$ and the number of non-zero entries $k$ at each node.

\section{Convergence Proof}

\begin{algorithm}[t!]
	\caption{\mml{} Quantized TopK SGD at a node $i$.}
	\label{algo:qtopk-sgd}
	\begin{algorithmic}
		\STATE {\bfseries Input:} Stochastic Gradient $\nabla G^i(\cdot)$ at node $i$
		\STATE {\bfseries Input:} value $K$, learning rate $\alpha$
		\STATE  Initialize $v_0 = \epsilon_0^i = \vec{0}$
		\FOR{ each step $t \geq 1$}
		\STATE ${acc}_t^i \gets \epsilon_{t - 1}^i + \alpha \nabla G^i_t(v_{t - 1}) $ \COMMENT{accumulate error into a locally generated gradient}
		\STATE $\epsilon_t^i \gets {acc}_t^i -  \mathsf{TopK}( {acc}_t^i ) $ \COMMENT{update the error}
		\STATE $g_t^i \gets \mathsf{allreduce}( Q( \mathsf{TopK}( {acc}_t^i ) ), \mathsf{SUM}
		)$ \COMMENT{ sum (sparse) contribution from all nodes }
		\STATE  $v_{t}^i \gets v_{t-1}^i - g_t^i$  \COMMENT{ apply the update }
		\ENDFOR
	\end{algorithmic}
\end{algorithm}

\paragraph{Preliminaries.} 
We consider a setting where the Quantized TopK SGD algorithm is minimizing a (non-convex) function $f$, following the SGD iteration with decreasing step sizes $\alpha_t$. 
We assume that $f$ is $L$-Lipschitz smooth, that is:

\begin{definition}
	\label{def:smooth-sc}
	For any differentiable function $f \colon \mathbb{R}^d \to \mathbb{R}$, 
	$f$ is $L$-Lipschitz smooth if
	$\forall x,y\in \mathbb{R}^d$, $\|\nabla f(x) - \nabla f(y)\|\leq L \|x-y\|$.
\end{definition}

Further, we assume that the second moment of the average of $P$ stochastic gradients with respect to any choice of parameter values is bounded, i.e.:
\begin{equation}
\label{eqnConvexNoise}
E \left[ {\normsmall{\frac{1}{P}\sum_{p=1}^P \tilde G^p(x)}^2} \right] \le M^2, \forall x\in\mathbb{R}^n
\end{equation}
where $\tilde G^1(x), \ldots, \tilde G^P(x)$ are $P$ independent stochastic gradients (at each node).

Moreover, to simplify the exposition we will slightly overload notation and re-define $M$ to satisfy the bound
\begin{equation}
\label{eqnConvexNoise}
E \left[ Q( {\normsmall{\frac{1}{P}\sum_{p=1}^P \tilde G^p(x)}^2} ) \right] \le M^2, \forall x\in\mathbb{R}^n, 
\end{equation}
where $Q$ is the quantization operator, and the expectation is taken over the randomness in the stochastic quantization, and over that of the stochastic sampling procedure. 
(Recall that the variance added by the stochastic quantization is upper bounded by $(1 + d^2/s)$, where $d$ is dimension and $s$ is the number of quantization points~\cite{alistarh2016qsgd}. 
This quantity can be directly multiplied into the standard second moment bound $M$ to obtain the above bound.) 

\paragraph{Analysis Preliminaries.}
Define $\tilde{G}_t(v_t) = \frac{1}{P} \sum_{p = 1}^P \tilde{G}^p_t\left(v_t\right)$. In the following, it will be useful to track the following auxiliary random variable at each global step $t$: 
\begin{equation}
\label{def:x}
x_{t+1} = x_t - \frac{1}{P} \sum_{p = 1}^P \alpha\tilde{G}^p_t\left(v_t\right) = x_t - \alpha\tilde{G}_t(v_t),
\end{equation}

\noindent where $x_0 = 0^n$. 
Intuitively, $x_t$ tracks all the gradients generated so far, without truncation or quantization. 
One of our first objectives will be to bound the difference between $x_t$ and $v_t$ at each time step $t$. 
Define: 
\begin{equation}
\label{def:eps}
\epsilon_t = \frac{1}{P} \sum_{p = 1}^P \epsilon_t^p. 
\end{equation}

The variable $x_t$ is set up such that, by induction on $t$, one can prove that, for any time $t \geq 0$, 
\begin{equation}
\label{eq:eps}
E_Q [ v_t - x_{t} ] = \epsilon_t, 
\end{equation}
where the expectation is taken over the randomness in the quantization operator $Q$. 

\paragraph{Commutativity of sum and TopK.} 
Following~\cite{alistarh2018convergence}, we will make the following analytic assumption about the proportion of the generated gradients which is preserved by the sum of TopK operators across all nodes:

\begin{assumption}
	\label{assumption1}
	There exists a (small) constant $\xi$ such that, for every iteration $t \geq 0$, we have:
	\begin{eqnarray*}
		\label{eq:assumption}
		\left\Vert \mathsf{TopK}\left( \frac{1}{P} \sum_{p = 1}^P \left( \alpha\tilde G_t^p ( v_t) + \epsilon_t^p \right) \right) - 
		\sum_{p = 1}^P \frac{1}{P} \mathsf{TopK}\left( \alpha\tilde G_t^p ( v_t) + \epsilon_t^p \right) \right\Vert \\ \leq \xi \|  \alpha\tilde G_t(v_t) \|.  
	\end{eqnarray*}
\end{assumption}

\paragraph{Analysis.} As is standard in non-convex settings~\cite{liu2015asynchronous}, we settle for a weaker notion of convergence, namely:
$$\min_{t\in \{1,\ldots,T\}} E {\|\nabla f\left(v_t\right)\|^2} \stackrel{T \rightarrow \infty}{\longrightarrow} \, 0,$$

\noindent that is, the algorithm converges ergodically to a point where gradients are $0$. Our strategy will be to leverage the  bound on the difference between the ``real'' model $x_t$ and the view $v_t$ observed at iteration $t$ to bound the expected value of $f(v_{t})$, which in turn will allow us to bound
$$
\frac{1}{\sum_{t=1}^{T}\alpha_t}\sum_{t=1}^{T}\alpha_t E{\|\nabla f\left(v_t\right)\|^2},
$$
\noindent where the parameters $\alpha_t$ are appropriately chosen \emph{decreasing} learning rate parameters. We start from:


\begin{lemma}
	\label{lemma_for_noncenvex}
	For any time $t\geq 1$:
	\begin{eqnarray*} 
		E_Q [ \|v_t - x_t\|^2 ] \leq \left(1 + \frac{\xi}{P\gamma}\right)^2\sum_{k=1}^t \left(2\gamma^2\right)^{k}\|x_{t-k+1} - x_{t-k}\|^2.
	\end{eqnarray*} 
\end{lemma}

\noindent We can then prove the following: 

\begin{theorem}
	\label{theorem_nonconvex}
	Consider Algorithm 1 for minimising a function $f$ that satisfies the  assumptions in this section. Suppose that the learning rate sequence and $K$ are chosen so that for any time $t>0$:
	\begin{equation}
	\label{non_convex_assumption_main}
	\sum_{k=1}^{t}\left(2\gamma^2\right)^k\frac{\alpha_{t-k}^2}{\alpha_{t}} \leq D
	\end{equation}
	for some constant $D>0$. Then, after running Algorithm 1 for $T$ steps:
	\begin{equation}
	\begin{split}
	\label{result_non_convex}
	\min_{t\in\{1, 2, \ldots, T\}} \mathbb{E}  \, [ \|\nabla f\left(x_t\right)\|^2 ] \leq  \\ \frac{1}{\sum_{t=1}^{T}\alpha_t}\sum_{t=1}^{T}\alpha_t E{\|\nabla f\left(v_t\right)\|^2} \leq \frac{4\left(f\left(x_0\right) - f\left(x^{*}\right)\right)}{\sum_{t=1}^{T}\alpha_t} \\  + \frac{\left(2LM^2 + 4L^2M^2\left(1 + \frac{\xi}{P\gamma}\right)^2D\right)\sum_{t=1}^{T}\alpha_t^2}{\sum_{t=1}^{T}\alpha_t}.
	\end{split}
	\end{equation}
\end{theorem}

\begin{table*}[th]
	\centering
	\begin{tabular}{ @{} l | r | r | r @{} }
		\toprule
		Name & \# Model & \# Global Batch Size & Number of Epochs \\
		\midrule 
		CIFAR-10 & ResNet-110 & 256 & 160 \\
		\midrule
		ImageNet-1K & 4xResNet 18 and 34 &  512 & 70 \\
		\midrule
		ATIS~\cite{hemphill1990atis} & LSTM  & 560 & 20  \\
		\midrule
		Hansards~\cite{hansards} & LSTM & 256   & 20 \\
		\midrule
		ASR (proprietary) & LSTM & 512   & 20 \\
		\bottomrule
	\end{tabular}
	
	\caption{Hyperparameters for our DNN training examples.  }
	\label{tbl:hyperparameters}
	\vspace{-0.5cm}
\end{table*}

Given the above updated definitions and preliminaries (notably the updated definition of the second moment bound $M$ and of the per-iteration bounds between $x_t$ and $v_t$), the proof will proceed identically to that of~\cite{alistarh2018convergence}.

\section{Hyperparameter Values used for DNN Experiments}

In this section, we detail the hyperparameter values used for our DNN training experiments. 
For all the experiments, we closely followed the standard hyperparameter values and learning rate schedules given in the CNTK repository, which are optimized for sequential execution. 
The exact BrainScript files we used (detailing the exact parameters) are therefore similar to those presented found at~\cite{CNTKEX}. For completeness, we detail the main values in Table~\ref{tbl:hyperparameters}. 
For the wide residual networks, we used the same learning rate schedule as for standard residual networks on ImageNet-1K, as given in~\cite{CNTKEX}: 
we start with a learning rate of $1$, which is divided by $10$ at $30$ and $60$ epochs. 
For the proprietary ASR model, we used an initial learning rate of $0.001$, and a global batch size of $512$ samples per iteration.

	
\end{document}